\documentclass[%
 reprint,
superscriptaddress,
 amsmath,amssymb,
 prx,
]{revtex4-2}

\usepackage{graphicx}% Include figure files
\usepackage{dcolumn}% Align table columns on decimal point
\usepackage{bm}% bold math
\usepackage{epsfig}
\usepackage{subcaption}
\usepackage{xcolor}

\usepackage{ragged2e}

\usepackage{hyperref}
\hypersetup{
    colorlinks=true,
    linkcolor=blue,
    filecolor=magenta, 
    }
\newcommand{\ah}[0]{\ensuremath{a}}
\newcommand{\ad}[0]{\ensuremath{a^\dagger}}

\newcommand{\Ham}[0]{\ensuremath{H}}

\newcommand{\imag}[0]{\ensuremath{\mathrm{i}}}
\newcommand{\omegaq}[0]{\ensuremath{\omega_\mathrm{q}}}
\newcommand{\omegad}[0]{\ensuremath{\omega_\mathrm{d}}}

\begin{document}

\preprint{APS/123-QED}

\title{Reducing leakage of single-qubit gates for superconducting quantum processors using analytical control pulse envelopes}

\author{Eric Hyyppä}
\email{Corresponding author: eric@meetiqm.com}
\affiliation{%
 IQM Quantum Computers, Keilaranta 19, 02150 Espoo, Finland
}%

\author{Antti Vepsäläinen}%
\affiliation{%
 IQM Quantum Computers, Keilaranta 19, 02150 Espoo, Finland
}%

\author{Miha Papi\v{c}}
\affiliation{%
 IQM Quantum Computers, Georg-Brauchle-Ring 23-25, 80992 Munich, Germany
}%
\affiliation{%
 Department of Physics and Arnold Sommerfeld Center for Theoretical Physics, Ludwig-Maximilians-Universität München, Theresienstr. 37, 80333 Munich, Germany
}%

\author{Chun Fai Chan}
\affiliation{%
 IQM Quantum Computers, Keilaranta 19, 02150 Espoo, Finland
}%

\author{Sinan Inel}
\affiliation{%
 IQM Quantum Computers, Keilaranta 19, 02150 Espoo, Finland
}

\author{Alessandro Landra}
\affiliation{%
 IQM Quantum Computers, Keilaranta 19, 02150 Espoo, Finland
}%

\author{Wei Liu}
\affiliation{%
 IQM Quantum Computers, Keilaranta 19, 02150 Espoo, Finland
}%

\author{Jürgen Luus}
\affiliation{%
 IQM Quantum Computers, Keilaranta 19, 02150 Espoo, Finland
}%

\author{Fabian Marxer}
\affiliation{%
 IQM Quantum Computers, Keilaranta 19, 02150 Espoo, Finland
}%

\author{Caspar Ockeloen-Korppi}
\affiliation{%
 IQM Quantum Computers, Keilaranta 19, 02150 Espoo, Finland
}%

\author{Sebastian Orbell}
\affiliation{%
 IQM Quantum Computers, Keilaranta 19, 02150 Espoo, Finland
}
\affiliation{%
 Department of Materials, University of Oxford, Oxford, UK
}%

\author{Brian Tarasinski}
\affiliation{%
 IQM Quantum Computers, Keilaranta 19, 02150 Espoo, Finland
}%

\author{Johannes Heinsoo}
\affiliation{%
 IQM Quantum Computers, Keilaranta 19, 02150 Espoo, Finland
}%

\date{August 23, 2024}
%\date{April 15, 2025}

\begin{abstract}
Improving the speed and fidelity of quantum logic gates is essential to reach quantum advantage with future quantum computers. However, fast logic gates lead to increased leakage errors in superconducting quantum processors based on qubits with low anharmonicity, such as transmons. 
To reduce leakage errors, we propose and experimentally demonstrate two new analytical methods, Fourier ansatz spectrum tuning derivative removal
by adiabatic gate (FAST DRAG) and higher-derivative (HD) DRAG, both of which enable shaping single-qubit control pulses in the frequency domain to achieve stronger suppression of 
leakage transitions compared to previously demonstrated pulse shapes.
Using the new methods to suppress the $ef$-transition of a transmon qubit with an anharmonicity of -212 MHz, we implement $R_X(\pi/2)$-gates achieving a leakage error below $3.0 \times 10^{-5}$ down to a gate duration of 6.25 ns without the need for iterative closed-loop optimization. The obtained leakage error represents a 20-fold reduction in leakage compared to a conventional Cosine DRAG pulse. 
Employing the FAST DRAG method, we further achieve an error per gate of $(1.56 \pm 0.07)\times 10^{-4}$ at a 7.9-ns gate duration, outperforming conventional pulse shapes both in terms of error and gate speed. Furthermore, we study error-amplifying measurements for the characterization of temporal microwave control pulse distortions, and demonstrate that non-Markovian coherent errors caused by such distortions may be a significant source of error for sub-10-ns single-qubit gates unless corrected using predistortion.

%\begin{description}
%\end{description}
\end{abstract}

\maketitle

%\tableofcontents

\section{\label{sec:intro} Introduction}

Superconducting qubits provide a promising platform for realizing large-scale quantum computers as they enable high-fidelity quantum logic gate operations in the nanosecond regime \cite{google2023suppressing, mckay2023benchmarking, li2023error, marxer2023long, sung2021realization, Ding2023_FTF, werninghaus2021leakage, rower2024suppressing}, potentially resulting in significantly faster computations compared to other state-of-the-art quantum computing platforms, such as neutral atoms \cite{shi2022quantum, bluvstein2023logical, evered2023high} or trapped ions \cite{harty2014high, gaebler2016high}. Improving the gate speed further would not only increase the clock speed of the quantum computer but it would also reduce the incoherent gate errors caused by decoherence, thus enabling even higher fidelities. Currently, the large-scale superconducting quantum processing units (QPU) are based on transmon qubits which suffer from frequency crowding and a low anharmonicity typically in the range of $|\alpha/(2\pi)| \in [150, 300]$ MHz \cite{google2023suppressing, mckay2023benchmarking, krinner2022realizing}. Thus, fast logic gates cause the frequency spectra of the control pulses to overlap with transitions out of the computational subspace -- such as the $ef$-transition -- resulting in leakage errors \cite{motzoi2009simple, motzoi2013improving, chen2016measuring, mckay2017efficient}. 
Leakage is particularly harmful for quantum error correction applications \cite{aliferis2005fault, fowler2013coping, suchara2015leakage} in transmon-based processors due to the generation of non-local errors as a result of leakage being propagated by entangling two-qubit gates~\cite{mcewen2021removing}.

There are two main strategies to mitigate leakage errors in superconducting qubits. The first strategy aims to prevent the occurrence of leakage in the first place, e.g., using pulse shape engineering \cite{motzoi2009simple, motzoi2013improving, chen2016measuring, mckay2017efficient, chow2010optimized, gambetta2011analytic, vesterinen2014mitigating, martinis2014fast, theis2016simultaneous, theis2018counteracting, werninghaus2021leakage, google2023suppressing, li2024experimental, ribeiro2017systematic, setiawan2021analytic}, whereas the second strategy relies on leakage reduction units converting already occurred leakage errors into Pauli errors \cite{mcewen2021removing, lacroix2023fast, marques2023all}. In this work, we focus on the first strategy that enables the reduction of the overall error for fast logic operations. 
 Leakage errors caused by single-qubit control pulses are often suppressed using the derivative removal by adiabatic gate (DRAG) method that sets the quadrature envelope of a control pulse equal to a scaled derivative of its in-phase envelope \cite{motzoi2009simple, chen2016measuring, mckay2017efficient}. This effectively suppresses the frequency spectrum of the control pulse at a specific frequency, such as the $ef$-transition frequency, reducing leakage related to the suppressed transition. Additionally, the leakage error can be reduced by the choice of the control pulse envelopes, for example, using the Slepian pulse shape \cite{martinis2014fast, sung2021realization, marxer2023long} that is commonly applied in two-qubit gates to minimize the spectral energy above a given cutoff frequency.
Furthermore, optimal control methods have been shown to reduce leakage errors of fast gates. For example, the lowest leakage error for fast transmon-based single-qubit gates has been achieved using experimental closed-loop optimization of piecewise constant $R_X(\pi/2)$ control pulses with 23 parameters, which  resulted in a leakage rate per  Clifford gate of $4.4 \times 10^{-4}$ at a pulse duration of $t_\mathrm{g} = 4.16$ ns ($t_\mathrm{g} |\alpha| /(2\pi) \approx 1.31$) \cite{werninghaus2021leakage}. 
However, such closed-loop optimization methods may require a large number of iterations, leading to increased calibration time \cite{werninghaus2021leakage}. Open-loop optimal control techniques may provide significantly faster convergence by optimizing the control pulses in a simulation \cite{khaneja2005optimal, caneva2011chopped, rach2015dressing, machnes2018tunable, niu2019universal, song2022optimizing}, but the obtained pulse shapes may not transfer well to an experimental implementation if the assumed model Hamiltonian does not capture all the relevant details of the experimental system, such as electronics limitations, pulse distortions or parasitic interactions.

Two-qubit gate fidelities in superconducting quantum processors have recently improved considerably \cite{sung2021realization, stehlik2021tunable, marxer2023long, Ding2023_FTF, zhang2023tunable, li2024realization}, and hence, further improvements are needed for single-qubit gates that typically appear in quantum circuits in higher numbers compared to two-qubit gates \cite{barends2014superconducting,stehlik2021tunable, krinner2022realizing}. 
In this work, we demonstrate two new methods for engineering single-qubit gate pulse shapes
that enable a more flexible control of the frequency spectrum resulting in significantly reduced leakage errors 
for fast gates without the need for closed-loop optimization. Applying the new methods to implement single-qubit gates on a transmon with an anharmonicity of $\alpha/(2\pi)=-212$ MHz, we experimentally achieve an average leakage error per $R_X(\pi/2)$-gate below $3 \times 10^{-5}$ down to a gate duration of $t_\mathrm{g}=6.25$ ns ($t_\mathrm{g}|\alpha|/(2\pi) \approx 1.33$) approaching the speed limit of single-qubit gates $t_\mathrm{g} \sim 2\pi/|\alpha|$ \cite{theis2016simultaneous, zhu2021quantum}. At such short gate durations, the proposed methods reduce the leakage error by a factor of 20 compared to DRAG with a conventional cosine envelope and by an order of magnitude compared to the previous lowest leakage rate at a comparable gate duration \cite{werninghaus2021leakage}. 
Importantly, one of the novel approaches presented here, Fourier ansatz spectrum tuning (FAST) DRAG, enables the implementation of a 7.9-ns $R_X(\pi/2)$-gate with a gate error of $(1.56 \pm 0.07) \times 10^{-4}$, reducing the gate error and improving the gate speed over the conventional pulse shapes. In addition to low errors, the proposed methods also reduce the required peak power of sub-10-ns single-qubit gates compared to conventional pulse shapes. Furthermore, we study error mechanisms relevant for fast single-qubit gates with fidelities approaching 99.99\%, and find that non-Markovian coherent errors caused by microwave control pulse distortions \cite{gustavsson2013improving} may be a major error source in addition to incoherent errors and leakage. 
We characterize the microwave pulse distortions in our experimental system using simple error amplification experiments and show that the gate fidelity can be improved by predistorting the microwave control pulses.

\section{\label{sec: control pulses} Control pulses for suppressing leakage errors}

\subsection{\label{sec: DRAG} Introduction to single-qubit gates and DRAG}

\begin{figure*}
\centering
    \includegraphics[width=0.95\textwidth]{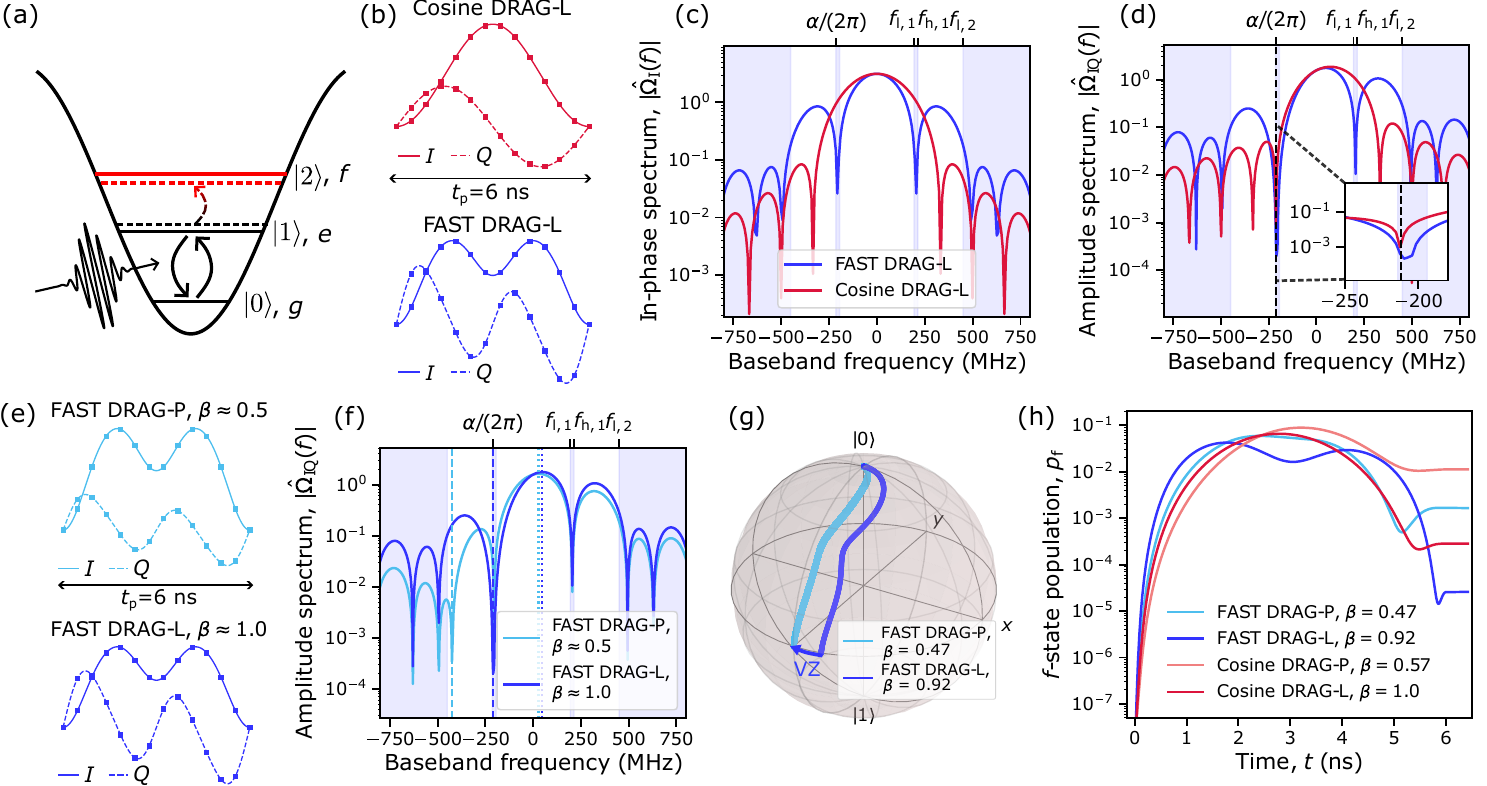}
\caption{ \label{fig: intro and FAST} \justifying  \textbf{Fourier anzatz spectrum tuning for fast low-leakage single-qubit gates.} 
(a) Lowest energy levels of a transmon qubit with a control pulse inducing state transfer (arrows), leakage errors (red) and AC Stark shift (dashed lines).
(b) Envelopes of the in-phase ($I$, solid) and quadrature ($Q$, dashed) components for a conventional cosine DRAG pulse (red here and later) and the proposed FAST DRAG pulse (blue here and later) sampled (square markers) for our AWG. 
(c) Amplitude spectrum $|\hat{\Omega}_\mathrm{I}(f)|$ of the in-phase envelope $\Omega_\mathrm{I}$ for the same pulses, with the blue regions illustrating the frequency intervals, across which the spectral energy of the in-phase envelope is suppressed for the FAST DRAG pulse. 
(d) Amplitude spectrum $|\hat{\Omega}_\mathrm{IQ}(f)|$ of the complex $IQ$ envelope $\Omega_\mathrm{IQ}$ for the same pulses using DRAG to suppress the spectrum at the qubit anharmonicity $\alpha/(2\pi)$ (dashed line). The inset shows a close-up around $\alpha/(2\pi)$. 
(e) In-phase ($I$) and quadrature ($Q$) envelopes for two variants of FAST DRAG with $\beta \approx 0.5$ (light blue) resulting in minimal phase errors (FAST DRAG-P) and  $\beta \approx 1.0$ (dark blue) resulting in minimal leakage (FAST DRAG-L).
(f) Amplitude spectrum $|\hat{\Omega}_\mathrm{IQ}(f)|$ of the complex $IQ$ envelope for the same pulses, with frequencies suppressed by DRAG (dashed lines), and frequencies corresponding to maximal spectral energy density (dotted lines).
(g) Simulated trajectory of the qubit state during a $R_X(\pi/2)$-gate for the same pulses, with a virtual Z-rotation (blue arrow) needed with the FAST DRAG-L pulse. 
(h) Simulated population leakage of a $R_X(\pi/2)$-gate starting from $|1\rangle$ for the same pulses together with Cosine DRAG-P and Cosine DRAG-L. 
}
\end{figure*}

Single-qubit rotations of superconducting qubits are typically implemented with nearly resonant microwave drive pulses as illustrated in Fig.~\ref{fig: intro and FAST}(a) with the drive pulses applied to a drive line which is either capacitively or inductively coupled to the qubit. In the frame rotating at the microwave drive frequency $f_\mathrm{d}=\omega_\mathrm{d}/(2\pi)$, the Hamiltonian of a driven non-linear oscillator, such as a transmon qubit, can be written as \cite{motzoi2009simple}
\begin{equation}
    H_\mathrm{R} =\hbar  \sum_j \left( \delta_j |j\rangle \langle j| + \frac{\Omega_\mathrm{I}(t)}{2} \sigma_{j-1,j}^x +\frac{\Omega_\mathrm{Q}(t)}{2} \sigma_{j-1,j}^y  \right ), \label{eq: Ham driven qb with RWA}
\end{equation}
where $\delta_j = \omega_j - j\omega_\mathrm{d}$ is the detuning corresponding to the $j$th energy level with energy $\hbar \omega_j$, $\Omega_\mathrm{I}(t)$ and $\Omega_\mathrm{Q}(t)$ are the in-phase and quadrature components of the complex envelope $\Omega_\mathrm{IQ}(t) = \Omega_\mathrm{I}(t) - \imag \Omega_\mathrm{Q}(t)$ related to the control pulse $\Omega_\mathrm{RF}(t) = \Omega_\mathrm{I}(t)\cos(\omega_\mathrm{d}t) + \Omega_\mathrm{Q}(t)\sin(\omega_\mathrm{d}t)$ with a pulse duration $t_\mathrm{p}$, and the Pauli-like operators are given by $\sigma_{j-1,j}^x= \lambda_j (|j\rangle \langle j-1| +|j-1\rangle \langle j|)$ and $\sigma_{j-1,j}^y= \mathrm{i} \lambda_j (|j\rangle \langle j-1| -|j-1\rangle \langle j|)$. For a resonantly-driven transmon qubit, the drive frequency $\omega_\mathrm{d}$ matches the qubit frequency $\omegaq = \omega_1 - \omega_0$, and the state space can be approximately truncated to the three lowest levels with $\delta_1=0$, $\delta_2 = \omega_2-2\omega_1=\alpha$, $\lambda_1=1$, and $\lambda_2\approx\sqrt{2}$.  From Eq.~\eqref{eq: Ham driven qb with RWA}, we observe that $\Omega_\mathrm{I}$ ($\Omega_\mathrm{Q}$) induces rotations around the $x$-axis ($y$-axis) in the computational subspace with the angle of rotation controlled by the integral of the envelope. 

% Paragraph on error sources
For superconducting qubits, errors during single-qubit gates are mainly caused by  decoherence and the coupling of the qubit drive to state transitions other than the target transition, which induces phase errors and leakage out of the computational subspace if left uncorrected  \cite{chow2010optimized, lucero2010reduced, chen2016measuring, mckay2017efficient, papic2023fast, lazar2023improving}. Phase errors arise if the phase of the microwave drive evolves at a different rate compared to the phase of the qubit. Even though the drive frequency would be equal to the qubit frequency during idling, phase errors may arise during gates due to an AC Stark shift of the qubit frequency $\delta_{\rm AC}(t) \approx -\lambda_2^2\Omega_\mathrm{I}(t)^2/(4\alpha)$ resulting from a coupling between the higher energy states and the qubit drive, see Fig.~\ref{fig: intro and FAST}(a). 
Leakage errors during fast single-qubit gates are primarily caused by the frequency spectrum of the control pulse overlapping with leakage transitions. 
Thus, leakage can be reduced by engineering control pulses with a low spectral amplitude $|\hat{\Omega}_\mathrm{RF}(f)| = |\int \Omega_\mathrm{RF}(t) \exp(-\mathrm{i}2\pi f t) \mathrm{d}t|$ at undesired leakage transitions \cite{motzoi2009simple, motzoi2013improving}.

% Paragraph on DRAG-P solution 
Conventionally, single-qubit control pulses are shaped with DRAG to reduce either phase errors -- which we here refer to as DRAG-P -- or both phase and leakage errors, which we refer to as DRAG-L \cite{chen2016measuring, mckay2017efficient}. For both of the DRAG variants, the quadrature envelope is given by $\Omega_\mathrm{Q}(t) = -\beta\dot{\Omega}_I(t)/\alpha$, where $\beta$ denotes the DRAG coefficient, and $\Omega_\mathrm{I}(t)$ is  the in-phase pulse envelope, which can be, e.g., a Gaussian with $\Omega_\mathrm{I}(t)=\mathrm{exp}[-(t-t_\mathrm{p}/2)^2/(2\sigma^2)] -\mathrm{exp}[-t_\mathrm{p}^2/(8\sigma^2)]$ or a raised cosine with  $\Omega_\mathrm{I}(t) = [1 - \cos(2\pi t/t_\mathrm{p})]/2$, as shown in Fig.~\ref{fig: intro and FAST}(b). 
In the DRAG-P variant, the DRAG coefficient $\beta$ is tuned to cancel the AC Stark shift during the gate, which for a transmon occurs at $\beta \approx 0.5$ \cite{lucero2010reduced, gambetta2011analytic}. For sufficiently slow gates, the leakage errors can be ignored.

% Paragraph on DRAG-L solution 
In the second variant, DRAG-L, the DRAG coefficient $\beta$ is tuned such that the spectrum of the control pulse has a zero crossing in the vicinity of the $ef$-transition frequency, which  occurs at $\beta\approx 1.0$. This spectral suppression leads to a lower leakage error, thus potentially enabling faster and more accurate gates compared to DRAG-P. Despite lower errors provided by DRAG-L, the DRAG-P variant has been more commonly used in literature due to easier calibration and the dominance of phase errors over leakage in early superconducting qubits~\cite{lucero2010reduced,chow2010optimized, reed2013entanglement, bengtsson2020quantum, hyyppa2022unimon}. Namely, the choice of $\beta\approx 1.0$ is not optimal for eliminating the AC Stark shift, and thus phase errors need to be corrected with another approach for the DRAG-L variant, such as by detuning the drive frequency during the gate~\cite{chen2016measuring} or by applying virtual Z-rotations, i.e., updating the phase of the microwave drive, before and after each gate~\cite{mckay2017efficient}. If the phase errors are accounted for by tuning the drive frequency during the gate, the optimal values of the drive detuning and $\beta$ are coupled, which complicates calibration. On the other hand, virtual Z-rotations are straightforward to use with gate sets based on $R_X(\pi/2)$-rotations and two-qubit gates commuting with $\hat{\sigma}_z$. However, the implementation of a $R_X(\pi)$-gate requires two separate $R_X(\pi/2)$-gates as the qubit state cannot be evolved from pole to pole without applying virtual Z-rotations at some point between the poles. In principle, arbitrary-angle single-qubit gates can be decomposed with at most two $R_X(\pi/2)$-gates and three virtual Z-rotations \cite{mckay2017efficient}. In this work, we decompose the single-qubit Clifford group using $R_X(\pi/2)$-gates as explained in Sec.~\ref{sec: exp results} and use virtual Z-rotations to mitigate phase errors of DRAG-L pulses. %and a decompose single-qubit gates using  compatible  rotations that are especially convenient for gate sets based on $R_X(\pi/2)$ single-qubit rotations and two-qubit $CZ$ rotations. With this gate set, the implementation of a $R_X(\pi)$-gate, however, requires two separate $R_X(\pi/2)$-gates because the qubit state cannot be evolved from pole to pole  without applying virtual Z-rotations at some point between the poles. 

\subsection{\label{sec: FAST} Control pulses based on Fourier ansatz spectrum tuning}

For short microwave pulses or in the presence of multiple leakage transitions, it would be desirable to suppress the frequency spectrum across one or many frequency intervals of adjustable width, which is not possible with conventional DRAG pulses suppressing the spectrum across a single narrow frequency range. Importantly, the frequencies of the leakage transitions may shift during short gates due to AC Stark shifts or they may have finite widths induced by dephasing~\cite{lazar2023improving}. Thus, it may be beneficial to suppress the spectrum of the control pulse across wider frequency intervals around leakage transitions  to reduce the leakage errors of fast gates beyond the performance offered by conventional DRAG pulses. 

To overcome these issues, we present a novel pulse shaping method that parametrizes a control pulse envelope in terms of one or multiple configurable frequency intervals, across which the spectral energy is minimized to reduce leakage. We call the method Fourier ansatz spectrum tuning (FAST) since the spectral shaping is achieved using a control pulse envelope expressed as a Fourier cosine series
\begin{equation}
    \Omega_\mathrm{I}(t) = A \sum_{n=1}^N c_n g_n(t), \label{eq: FAST cos series}
\end{equation}
where $g_n(t)= [1 - \cos(2\pi n t /t_\mathrm{p})] \Pi(t/t_\mathrm{p} - 1/2)$ is the $n$th basis function parametrized to ensure continuity of the cosine series similar to Ref.~\cite{theis2016simultaneous}. Furthermore, $A$ is the overall amplitude, $N$ is the number of cosine terms, $\{c_n\}$ are the set of Fourier coefficients, $t_\mathrm{p}$ denotes the pulse duration, and $\Pi(x)$ is the rectangular window function between $x\in [-1/2, 1/2]$, which sets the pulse envelope to 0 outside of $[0, t_\mathrm{p}]$. 

To solve for the coefficients $\{c_n \}$, we analytically minimize the spectral energy of the control envelope across undesired frequency intervals as highlighted in Fig.~\ref{fig: intro and FAST}(c), while associating each interval with a  weight controlling the amount of suppression. The minimization problem can be written as
\begin{gather}
    \mathrm{min.}~\sum_{j=1}^k w_j \int_{f_{\mathrm{l}, j}}^{f_{\mathrm{h}, j}} |\hat{\Omega}_\mathrm{I}(f)|^2 \mathrm{d} f, \label{eq: FAST minimization} \\
    \mathrm{s.t.} \sum_{n=1}^N c_n t_\mathrm{p} = \theta, \label{eq: FAST constraint}
\end{gather}
where $k$ is the number of frequency intervals to suppress, the hat denotes Fourier transform, $w_j$ is the weight associated with the $j$th undesired frequency interval $[f_{\mathrm{l}, j},f_{\mathrm{h}, j}]$, and  the constraint fixes the rotation angle of the gate. Here, the frequencies are defined in the baseband, and the spectrum of the upconverted control pulse is suppressed across frequency intervals $\{f_\mathrm{d} \pm [f_{\mathrm{l}, j},f_{\mathrm{h}, j}]\}$ symmetrically located around the drive frequency $f_\mathrm{d}$. 
Importantly, the quadratic minimization problem with a linear constraint allows an analytical solution for the coefficient vector $\bm{c}\in \mathbb{R}^{N\times 1}$ for a given set of frequency intervals and weights by inverting the matrix equation
\begin{equation}
    \begin{bmatrix} \bm{A} + \bm{A}^T & -\bm{b} \\ \bm{b}^T &  0 \end{bmatrix} \begin{bmatrix} \bm{c}\\ \mu \end{bmatrix} = \begin{bmatrix} \bm{0} \\ \theta/t_\mathrm{p} \end{bmatrix}, \label{eq: c from matrix eq}
\end{equation}
where $\bm{A}_{nm} = \sum_{j=1}^k w_j \int_{f_{\mathrm{l}, j}}^{f_{\mathrm{h}, j}} \hat{g}_n(f)\hat{g}_m^*(f)\mathrm{d}f$, $\bm{b}=(1, \dots, 1)^T \in \mathbb{R}^{N \times 1}$, $\bm{0}=(0, \dots, 0)^T \in \mathbb{R}^{N \times 1}$, and $\mu$ is the Lagrangian multiplier. A more detailed derivation is provided in Appendix \ref{ap: FAST ap}. The FAST method can be viewed as an extension of the Slepian \cite{martinis2014fast}, and it parametrizes the control pulse in the frequency domain, while providing a mapping to the time-domain parameters $\{c_n\}$ using Eq.~\eqref{eq: c from matrix eq}. Alternatively, one can view $N$, $\{[f_{\mathrm{l}, j},f_{\mathrm{h}, j}]\}$, and $\{w_j \}$ as hyperparameters controlling the pulse shape. 

To implement fast, low-leakage single-qubit gates, we combine the FAST method with DRAG by using FAST to shape the in-phase envelope of the control pulse and applying DRAG to obtain the quadrature component. The resulting FAST DRAG pulse illustrated in Fig.~\ref{fig: intro and FAST}(b) has multiple benefits compared to using DRAG with traditional pulse envelopes, such as the raised cosine. First, the FAST DRAG pulse enables a stronger and wider spectral suppression of the $ef$-transition compared to conventional approaches as shown in Fig.~\ref{fig: intro and FAST}(d), which helps to mitigate leakage to the $f$-state. To achieve this, we apply the FAST method to suppress the spectrum of the in-phase component across a frequency interval $[f_{\mathrm{l}, 1},f_{\mathrm{h}, 1}]=[f_{\mathrm{l}, \mathrm{ef}},f_{\mathrm{h}, \mathrm{ef}}] \ni |\alpha|/(2\pi)$, and use DRAG-L with $\beta \approx 1$ to further suppress the Fourier transform $\hat{\Omega}_\mathrm{IQ}$ of the complex envelope  that is of the product form $\hat{\Omega}_\mathrm{IQ}(f) = [1 - 2\pi\beta f/\alpha]\times \hat{\Omega}_I(f)$. The strong suppression of $\hat{\Omega}_\mathrm{IQ}(f)$ around $\alpha/(2\pi)$ then leads to a low spectral amplitude $|\hat{\Omega}_\mathrm{RF}(f)|$ around the $ef$-transition frequency assuming $\omega_\mathrm{d} = \omega_\mathrm{q}$.
In addition to suppressing the signal around the $ef$-transition, we use a second frequency interval $[f_{\mathrm{l}, 2},f_{\mathrm{h}, 2}]$ to reduce the spectral energy above a given cutoff frequency $f_\mathrm{c} = f_{\mathrm{l}, 2}$. This favorably reduces the bandwidth of the control pulse and often indirectly lowers the peak amplitude, thus effectively regularizing the pulse shape. As a further benefit, it is relatively straightforward to calibrate FAST DRAG pulses since the parameters $A$, $\beta$, and $\{[f_{\mathrm{l}, j},f_{\mathrm{h}, j}]\}$ are mostly decoupled, and the locations of the suppressed frequency intervals $\{[f_{\mathrm{l}, j},f_{\mathrm{h}, j}]\}$ do not affect the effective drive frequency due to a symmetric suppression of the spectrum (see Fig.~\ref{fig: intro and FAST}(c)). 

In this work, we select the hyperparameters $\{N, f_{\mathrm{l}, \mathrm{ef}}, f_{\mathrm{h}, \mathrm{ef}}, f_\mathrm{c}, w_\mathrm{ef} = w_1/w_2 \}$ using physics-inspired heuristics based on the measured anharmonicity $\alpha/(2\pi)$ without optimizing the hyperparameters. 
In Appendix~\ref{ap: hyperparameter selection}, we provide general heuristic guidelines for the hyperparameter selection in systems with a single dominant leakage transition  and discuss an extension to systems with multiple leakage transitions important for simultaneous single-qubit gates. The heuristic hyperparameter selection is possible because the leakage error is relatively robust against moderate changes in the hyperparameters as discussed in Sec.~\ref{sec: exp results} and Appendix~\ref{ap: FAST param sweeps}. After the hyperparameter selection, the remaining single-qubit gate parameters can be calibrated similarly to conventional DRAG pulses, which renders the calibration of FAST DRAG pulses efficient and robust avoiding the challenges with closed-loop and open-loop optimal control methods mentioned in Sec.~\ref{sec:intro}.

When using FAST DRAG, we study both of the above-described DRAG-P and DRAG-L variants, the envelope pulses and spectra of which are illustrated in Figs.~\ref{fig: intro and FAST}(e) and (f). In our calibration, we set the drive frequency equal to the qubit frequency and use virtual Z-rotations to correct the phase errors caused by FAST DRAG-L pulses as illustrated in Fig.~\ref{fig: intro and FAST}(g).
According to simulations, FAST DRAG-L indeed achieves a lower simulated leakage error for a 6-ns $R_X(\pi/2)$-gate compared to FAST DRAG-P and the conventional cosine-based approaches, see Fig.~\ref{fig: intro and FAST}(h).

\subsection{\label{sec: higher-derivative DRAG} Control pulses based on higher-derivative (HD) extension of DRAG}

In this section, we present an alternative pulse shaping method for implementing fast, low-leakage single-qubit gates by extending the theoretical ideas of Ref.~\cite{motzoi2013improving}, in which a higher-derivative version of the DRAG method was theoretically proposed to mitigate multiple leakage transitions. In contrast to Ref.~\cite{motzoi2013improving}, we obtain $\Omega_\mathrm{Q}(t)$ from $\Omega_\mathrm{I}(t)$ using standard DRAG, decoupling the parameters of the in-phase and quadrature components, thus simplifying the calibration. Furthermore, we use the higher-order derivatives to achieve a strong spectral suppression of the dominant leakage transition, i.e., the $ef$-transition instead of suppressing multiple separate transitions as in Ref.~\cite{motzoi2013improving}. We also propose an ansatz based on a simple cosine series to ensure the continuity of the obtained control envelopes. 

In analogy to FAST DRAG, our version of the higher-derivative (HD) DRAG method (see Appendix~\ref{ap: HD DRAG ap}) minimizes the spectrum of the in-phase envelope around the qubit anharmonicity while additionally using DRAG to obtain the quadrature component and further suppressing the leakage. 
In this work, we consider a special case of the HD DRAG method containing derivatives up to the 2nd order in $\Omega_\mathrm{I}$ and up to the 3rd order in $\Omega_\mathrm{Q}$ as 
\begin{align}
    \Omega_\mathrm{I}(t) &= A [g(t) + \beta_2 \ddot{g}(t)], \label{eq: 3rd der DRAG I} \\
    \Omega_\mathrm{Q}(t) &= -\frac{A\beta}{\alpha} [\dot{g}(t) + \beta_2 \dddot{g}(t)], \label{eq: 3rd der DRAG Q}
\end{align}
where $g(t)$ denotes a basis envelope shape, and $\beta_2$ is a coefficient tuned to minimize the spectrum at the anharmonicity. Note that the orders of the derivative are not related to the order of DRAG as defined in Ref.~\cite{gambetta2011analytic}.

\begin{figure}
\centering
\includegraphics[width = 0.45\textwidth]{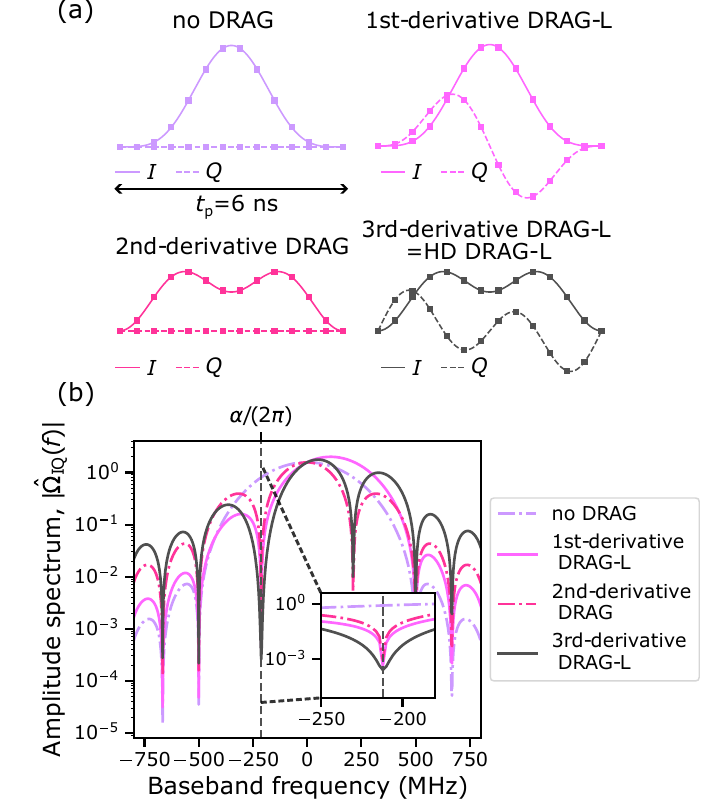}
    \caption{ \label{fig: higher-der DRAG} \justifying \textbf{Higher-derivative (HD) 
    DRAG.} (a) In-phase ($I$, solid) and quadrature ($Q$, dashed) envelopes of 6-ns HD DRAG pulses based on different maximum order of  derivatives sampled (square markers) for our AWG and compared to a pulse without DRAG.
    The pulse without DRAG sets  $\Omega_\mathrm{Q}=0$, the 1st-derivative DRAG pulse equals the conventional DRAG, the 2nd-derivative DRAG pulse sets $\Omega_\mathrm{I}$ according to Eq.~\eqref{eq: 3rd der DRAG I} with $\Omega_\mathrm{Q}=0$ similarly to the 2nd-derivative DRAG solution of Ref.~\cite{motzoi2013improving} , and the 3rd-derivative DRAG pulse is based on Eqs.~\eqref{eq: 3rd der DRAG I} and \eqref{eq: 3rd der DRAG Q}. 
    (b)  Amplitude spectrum $|\hat{\Omega}_\mathrm{IQ}(f)|$ of the complex $IQ$ envelope $\Omega_\mathrm{IQ}$ for the same pulses, with the qubit anharmonicity $\alpha/(2\pi)$ suppressed by DRAG shown with the dashed vertical line and the inset presenting a close-up around $\alpha/(2\pi)$.   }
\end{figure}

The Fourier transform of the complex envelope $\Omega_\mathrm{IQ}$ defined by Eqs.~\eqref{eq: 3rd der DRAG I} and \eqref{eq: 3rd der DRAG Q} is consequently given by $\hat{\Omega}_\mathrm{IQ}(f)=A \times [1 - \beta_2(2\pi f)^2]\times[1 - 2\pi\beta f/\alpha]\times \hat{g}(f)$, and the desired strong suppression at  $f=\alpha/(2\pi)$ is obtained by setting $\beta_2=1/\alpha^2$ and $\beta=1$ in line with DRAG-L. As $\hat{\Omega}_\mathrm{IQ}(f)$ factors as a product, the parameters $\beta$ and $\beta_2$ are decoupled simplifying calibration and allowing the use of DRAG-P to avoid phase errors if desired. This is not the case if the third derivative term  is left out from Eq.~\eqref{eq: 3rd der DRAG Q} like in the supplementary of Ref.~\cite{chen2016measuring}, where the minimization of leakage errors required a 2D sweep of $\{\beta, \beta_2\}$ and phase errors could not be avoided. 

Importantly, the basis function $g(t)$ and its three first derivatives should be continuous to ensure smooth control pulses, which we achieve by proposing the use of a short cosine series given by $g(t)=1-4/3\cos(2\pi t/t_\mathrm{p}) + 1/3\cos(4\pi t/t_\mathrm{p})$. The resulting envelopes of a third-derivative HD DRAG pulse in comparison to lower-derivative versions of DRAG are shown in Fig.~\ref{fig: higher-der DRAG}(a). After expressing $g(t)$ as a cosine series, this version of HD DRAG  is a special case of FAST DRAG with $N=2$ and a single suppressed frequency interval defined by $(f_{\mathrm{l}, \mathrm{ef}} + f_{\mathrm{h}, \mathrm{ef}}) /2 = |\alpha|/(2\pi)$ and $f_{\mathrm{h}, \mathrm{ef}} - f_{\mathrm{l}, \mathrm{ef}} \to 0$. 

Thus, HD DRAG and FAST DRAG have many common advantages since they share similar design criteria. Most importantly, HD DRAG enables strong spectral suppression of leakage transitions similarly to FAST DRAG  as illustrated in Fig.~\ref{fig: higher-der DRAG}(b). 
 Multiple leakage transitions can also be suppressed with HD DRAG by introducing higher-order derivatives as explained in Appendix~\ref{ap: HD DRAG ap}. Despite these similarities, there are also several key differences between the two methods. HD DRAG has less parameters than FAST DRAG, and the third-derivative HD DRAG studied in this work only introduces a single additional parameter $\beta_{2}$ determined by the measured anharmonicity $\alpha/(2\pi)$. However, HD DRAG does not allow straightforward control over the pulse bandwidth, the average power, or the exact strength and width of the spectral suppression unlike FAST DRAG, which offers more flexibility for the pulse design. Hence, we foresee that the flexibility of FAST DRAG is beneficial especially in scenarios with multiple leakage transitions, such as in simultaneous single-qubit gates as briefly discussed in Appendix~\ref{ap: hyperparameter selection}.

\section{\label{sec: exp results} Experimental implementation of fast, low-leakage single-qubit gates}

To experimentally demonstrate the proposed pulse shaping methods, we benchmark the error and leakage per gate for fast single-qubit gates implemented using FAST DRAG and HD DRAG and compare the results to conventional pulse shapes, such as Cosine DRAG. In this work, we utilize a two-qubit test device with a design and fabrication similar to Ref. \cite{marxer2023long}. We focus on one of the two flux-tunable qubits that is operated at its sweet-spot with a frequency of 4.417 GHz, an anharmonicity of -212 MHz, and average coherence times of about 37 us and 22 us for $T_1$ and $T_2^\mathrm{e}$ (Hahn echo), respectively. 
We implement single-qubit rotations by sending microwave pulses to a charge line coupled to the qubit through a targeted capacitance of 0.12 fF. The microwave pulses are generated by up-converting intermediate-frequency pulses from a Zurich Instruments HDAWG using an IQ-mixer, and then propagated to the sample attached to the cold-plate of a dilution refrigerator (Bluefors LD400) through coaxial lines with 67 dB room-temperature attenuation at 4.5 GHz, see Fig.~\ref{sfig: exp setup} in Appendix~\ref{ap: exp setup}. Since our gate decomposition is based on $R_X(\pi/2)$-gates, it is sufficient to calibrate the pulse amplitude for a single rotation angle $\theta = \pi/2$ allowing us to neglect drive non-linearities~\cite{lazuar2023calibration}.

\begin{figure*}
\centering
    \includegraphics[width = 0.92\textwidth]{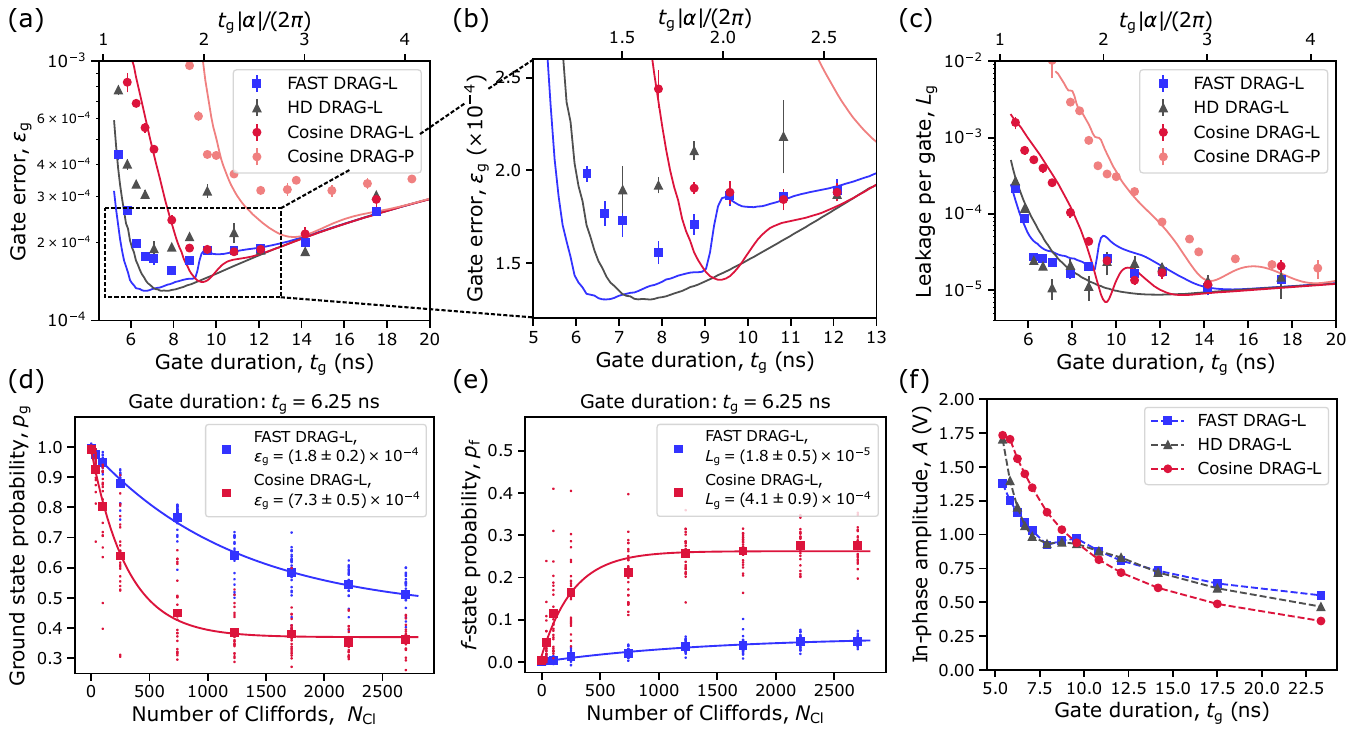}
    \caption{ \label{fig: leakage vs duration res} \textbf{Experimental benchmarking of fast low-leakage single-qubit gates.} \justifying (a) Experimental  error per gate $\varepsilon_\mathrm{g}$ from RB (markers) and simulated error for the $R_X(\pi/2)$-gate (solid lines) as functions of the gate duration for FAST DRAG-L (dark blue here and later), HD DRAG-L (dark grey here and later), Cosine DRAG-L (dark red here and later) and Cosine DRAG-P (light red here and later).   
    The error bars show 1$\sigma$ uncertainty of the mean based on 4-10 RB experiments.
    (b) Close-up to the gate error in panel (a) on the interval $t_\mathrm{g} \in [5, 13]$ ns. 
    (c) Experimental leakage per gate $L_\mathrm{g}$ from leakage RB (markers)  as a function of gate duration with simulated leakage error for the $R_X(\pi/2)$-gate (solid lines). 
    (d) Average ground state probability (square markers) as a function of  the number of Clifford gates in an RB experiment for FAST DRAG-L and Cosine DRAG-L with $t_\mathrm{g} = 6.25$ ns. The dots show results for 25 random Clifford sequences, while the solid lines show exponential fits estimating the error per gate $\varepsilon_\mathrm{g}$. 
    (e) Average $f$-state probability (square markers) as a function of the number of Clifford gates in the same RB experiment as in panel (d) with exponential fits estimating the leakage per gate $L_\mathrm{g}$ (solid lines). 
    (f) Amplitude $A$ of the in-phase envelope at the AWG output for the $R_X(\pi/2)$-gate as a function of the gate duration.
    }
\end{figure*}

To implement single-qubit gates using FAST DRAG, we suppress the spectrum of the in-phase component across a baseband frequency interval of $[f_{\mathrm{l}, 1},f_{\mathrm{h}, 1}] = [f_{\mathrm{l}, \mathrm{ef}},f_{\mathrm{h}, \mathrm{ef}}]=[194, 214]$ MHz covering the qubit anharmonicity, while using another frequency interval of $[f_{\mathrm{l}, 2},f_{\mathrm{h}, 2}] =[450, 1000]$ MHz effectively setting a cutoff frequency of $f_\mathrm{c}=450$ MHz. In addition, we use a weight ratio of $w_\mathrm{ef}=w_1/w_2=5$ ($w_\mathrm{ef}=100$) and set the number of cosine terms to $N=4$ ($N=5$) for FAST DRAG-L (DRAG-P). Here, a higher weight ratio is used for FAST DRAG-P to enable sufficient spectral suppression as the DRAG coefficient is reserved for correcting phase errors. 
The parameter values are chosen using physics-based heuristics and prior simulations to provide a low leakage rate across the studied gate durations  $t_\mathrm{g} \in [5, 20]$ ns. See Appendix~\ref{ap: hyperparameter selection} for general heuristic guidelines on the selection of these hyperparameters and Appendix~\ref{ap: FAST param sweeps} for parameter sweeps demonstrating that the leakage error stays low under moderate changes to the hyperparameter values.
For HD DRAG, we set $\beta_2 = 1 /\alpha^2$ to minimize the spectral energy density of the in-phase envelope at the anharmonicity. 
In all of the results of this section, we include a 0.41-ns delay between consecutive pulses as a part of the reported gate duration $t_\mathrm{g}$, i.e.,  $t_\mathrm{g} = t_\mathrm{p} + 0.41$ ns, and apply predistortion for the control pulse envelopes as discussed in more detail in Sec. \ref{sec: exp MW tails} and Appendix~\ref{ap: predistortion}.

To compare the error and leakage per gate for the proposed and conventional pulse shapes, we calibrate the single-qubit gate parameters $A$, $\beta$, $\omega_\mathrm{d}/(2\pi)$ and the virtual Z-rotation phase increment $\varphi_\mathrm{z}$ using sequences of error amplification experiments without any iterative closed-loop optimization as explained in Appendix \ref{ap: 1qb gate calib}, and benchmark the resulting gate performance using randomized benchmarking (RB) \cite{knill2008randomized, chow2009randomized, magesan2011scalable, epstein2014investigating} combined with 3-state discrimination (see Appendix~\ref{ap: 3state ro}) to further enable leakage RB analysis \cite{chen2016measuring,wood2018quantification}. 
The RB protocol enables estimating the average gate error by fitting an exponential model $p_\mathrm{g}(N_\mathrm{Cl}) = A_\mathrm{g} +B_\mathrm{g}p^{N_\mathrm{Cl}}$ to the ground state probability $p_\mathrm{g}$ measured as a function of the Clifford sequence length $N_\mathrm{Cl}$ and then extracting the average error per Clifford as $\varepsilon_\mathrm{Cl} = (1-p)/2$. Since we decompose the Clifford gates using on average  $N_\mathrm{g}\approx 2.21$ native gates from the gate set  $\{I, R_X(\pm\pi/2), R_Y(\pm\pi/2)\}$ \cite{epstein2014investigating}, we estimate the error per gate as $\varepsilon_\mathrm{g} = \varepsilon_\mathrm{Cl} / N_\mathrm{g}$ similarly to prior work using an equivalent Clifford decomposition \cite{werninghaus2021leakage, li2023error, mckay2017efficient}.
We further estimate the leakage error by fitting an exponential model $p_\mathrm{f}(N_\mathrm{Cl}) = A_\mathrm{f}  + B_\mathrm{f}\lambda_1^{N_\mathrm{Cl}}$ to the $f$-state population %as a function of $N_\mathrm{Cl}$ 
and then computing the leakage per gate as $L_\mathrm{g}=A_\mathrm{f}(1-\lambda_1)/N_\mathrm{g}$. 

From Fig.~\ref{fig: leakage vs duration res}(a) showing the simulated (see Appendix~\ref{ap: 1qb gate simulations}) and experimental gate error as a function of gate duration, we observe that Cosine DRAG-L calibrated to minimize leakage enables significantly lower gate errors compared to Cosine DRAG-P minimizing phase errors. 
For Cosine DRAG-P, the gate error begins to increase already below a gate duration of 14 ns, whereas the minimum gate error for Cosine DRAG-L reaches a value of $(1.84\pm 0.06)\times 10^{-4}$ at a gate duration of 10.8 ns. Importantly, the two novel pulse shapes based on FAST DRAG-L and HD DRAG-L enable a significantly reduced gate error compared to Cosine DRAG-L for fast gates with $t_\mathrm{g} \in [6, 9]$ ns. As illustrated in the close-up of Fig.~\ref{fig: leakage vs duration res}(b), FAST DRAG-L reaches a minimum average gate error of $(1.56 \pm 0.07) \times 10^{-4}$ at a gate duration of 7.9 ns, clearly outperforming the conventional Cosine DRAG-L pulse both in terms of gate error and duration. For FAST DRAG-L, the gate error stays at or below $2.0 \times 10^{-4}$ down to a gate duration of 6.25 ns. To reduce the effect of temporal fluctuations of $T_1$ in Figs.~\ref{fig: leakage vs duration res}(a)-(c), the gate error was measured for Cosine DRAG-L and FAST DRAG-L in an interleaved fashion and the full sweep was repeated on 3 separate days to acquire more statistics. The measurements for HD DRAG-L were conducted at a later date, during which $T_1$ was lower, explaining the slightly higher gate error compared to FAST DRAG-L. 

\begin{figure*}
\centering
    \includegraphics[width = 0.95\textwidth]{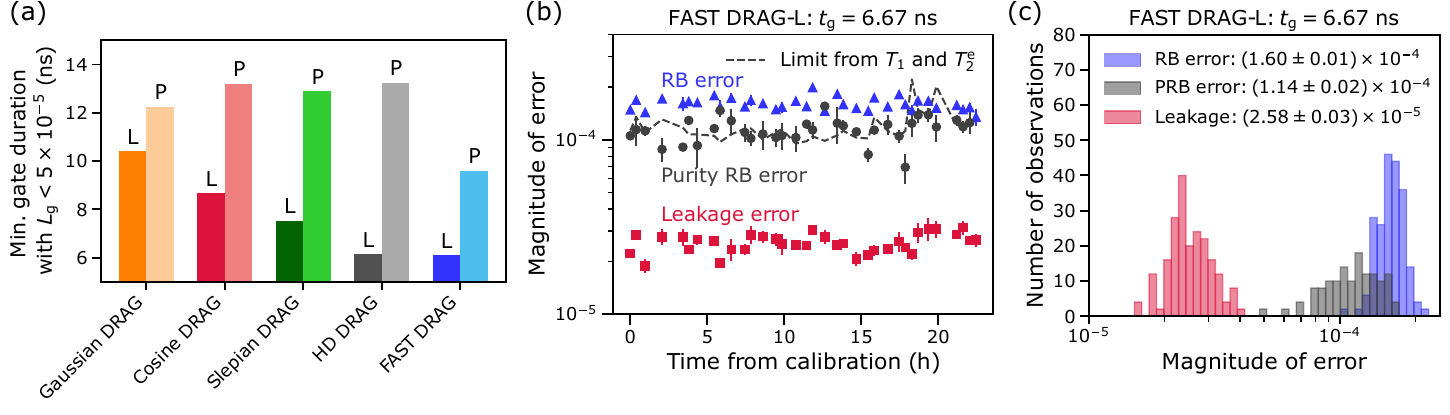}
    \caption{ \label{fig: stability and comparison} \justifying \textbf{Comparison of gate speed for different pulse shapes and temporal stability of FAST DRAG gates.} (a) Shortest gate duration with an experimental leakage error below $L_\mathrm{g}< 5 \times 10^{-5}$ for  conventional and proposed pulse shapes calibrated using DRAG-P (light color) or DRAG-L (dark color). For Gaussian DRAG, we use $\sigma = t_\mathrm{p} / 5$ and subtract away the discontinuity. For Slepian DRAG, we minimize the spectral energy above $f_\mathrm{c} = 185$ MHz. 
    See also Fig.~\ref{sfig: leakage for all pulses}(a)  for the duration-dependent leakage error of each pulse and Fig.~\ref{sfig: leakage for all pulses}(b) for an illustration of the pulse shapes, respectively.  (b) Experimental error per gate $\varepsilon_\mathrm{g}$ from RB (blue triangles), incoherent error per gate $\varepsilon_{\mathrm{g}, \mathrm{inc}}$ from purity RB (grey circles), and leakage per gate $L_\mathrm{g}$ from leakage RB (red squares) as a function of time since the calibration of a $R_X(\pi/2)$-gate using a FAST DRAG-L pulse with $t_\mathrm{g} = 6.67$ ns. %(including a 0.41 ns zero padding). 
    The markers and error bars are based on the mean and the standard error of mean estimated from three (leakage) RB and two purity RB experiments conducted at each point of time. The dashed grey line shows the coherence limit based on $T_1$ and $T_2^\mathrm{e}$.  (c) Histograms of RB gate error, purity RB error, and leakage error based on data in panel (b).    }
\end{figure*}

From the measured leakage error shown in Fig.~\ref{fig: leakage vs duration res}(c), we confirm that the error for short gates is dominated by leakage. We also find a good agreement between the measured leakage error and Master equation simulations based on a 4-level Duffing oscillator model even though the simulations ignore any potential non-idealities of the control electronics, such as finite bandwidth, dynamic range or sampling rate, see Appendix~\ref{ap: 1qb gate simulations} for more details.
Importantly, the experimental leakage per gate stays at or below $3.0 \times 10^{-5}$ down to a gate duration of 6.25 ns for both FAST DRAG-L and HD DRAG-L, providing up to a 20-fold reduction in leakage compared to Cosine DRAG-L. However, the leakage error for FAST DRAG-L showcases a shallow local maximum around $t_\mathrm{g}|\alpha|/(2\pi) \approx 2$ since our heuristics for the hyperparameter selection have been primarily designed to enable low leakage errors for fast gates $t_\mathrm{g}|\alpha|/(2\pi) \in [1, 2]$.
As shown in the example RB experiment of Figs.~\ref{fig: leakage vs duration res}(d) and (e), FAST DRAG-L with a gate duration of $t_\mathrm{g} = 6.25$ ns reduces the error and leakage per gate by factors of 4 and 23, respectively, compared to Cosine DRAG-L with equivalent duration.
The average leakage per gate of $(2.7\pm 0.3) \times 10^{-5}$ and the corresponding leakage per Clifford of $(6 \pm 1) \times 10^{-5}$ obtained using FAST DRAG-L at $t_\mathrm{g}=6.25$ ns, i.e., $t_\mathrm{g}|\alpha|/(2\pi) = 1.33$, are lower compared to the lowest value we have found in literature for such short pulses \cite{werninghaus2021leakage}. In Ref.~\cite{werninghaus2021leakage}, closed-loop optimization of a piecewise constant $R_X(\pi/2)$-pulse  resulted in an order of magnitude higher leakage per Clifford of $4.4 \times 10^{-4}$  using a transmon with an anharmonity of $-315$ MHz and a gate duration of 4.16~ns corresponding to a normalized gate duration of $t_\mathrm{g}|\alpha|/(2\pi) = 1.31$ practically equivalent to our work.

In comparison to conventional Cosine DRAG-L pulses, the novel pulse shapes do not require any higher peak amplitude, see Fig.~\ref{fig: leakage vs duration res}(f). In contrast, the required amplitudes of FAST DRAG-L and HD DRAG-L are significantly lower for gate durations below 9 ns due to a faster increase of voltage at the beginning and end of the pulse (see Figs.~\ref{fig: intro and FAST}(b) and \ref{fig: higher-der DRAG}(a)). For FAST DRAG-L and HD DRAG-L, the non-monotonic increase of peak amplitude when reducing gate duration is explained by the in-phase envelope changing from having a single maximum to having two maxima.

To compare the gate performance of the novel pulse shapes against other conventional pulses beyond cosine DRAG, we further experimentally benchmark the leakage rate for DRAG pulses based on Gaussian and Slepian in-phase envelopes. For the Gaussian envelope, we use $t_\mathrm{p}=5\sigma$ ns and subtract the discontinuity similar to, e.g., Ref.~\cite{gambetta2011analytic}. To construct the Slepian envelope, we use FAST pulse shaping with an effective cutoff frequency of $185~\mathrm{MHz} < \alpha/(2\pi)$. Further results obtained for Gaussian and Slepian envelopes with different parameter choices are provided in Appendices~\ref{ap: further leakage results} and \ref{ap: FAST param sweeps}. Based on measurements of leakage per gate as a function of gate duration in Appendix~\ref{ap: further leakage results}, we estimate an experimental speed limit for each of the control pulses by determining the shortest gate duration with a leakage per gate below $5\times 10^{-5}$. As shown in Fig.~\ref{fig: stability and comparison}(a), FAST DRAG-L and HD DRAG-L enable the fastest low-leakage gates, reaching the leakage threshold at a gate duration of approximately 6 ns. The corresponding speed limits are 7.5 ns, 8.7 ns, and 10.4 ns for Slepian DRAG-L, Cosine DRAG-L, and Gaussian DRAG-L. When calibrating DRAG to minimize phase errors, the corresponding speed limit is 12-13 ns for all the other pulses than FAST DRAG-P, for which the leakage rate stays below the threshold down to a gate duration of 9.6~ns thanks to a strong spectral suppression around the $ef$-transition provided by FAST pulse shaping.

We further study the temporal stability of gate errors and leakage for FAST DRAG-L pulses by calibrating a 6.67-ns $R_X(\pi/2)$-gate and subsequently conducting repeated measurements of RB, leakage RB, and purity RB over a time period of 23 hours. The purity RB experiment incorporates state tomography characterization after each random Clifford sequence to measure the average purity of the final state as a function of the Clifford sequence length, which enables estimating the incoherent error per gate as $\varepsilon_{\mathrm{g}, \mathrm{inc}} = 1/2\times (1 - \sqrt{u})/N_\mathrm{g}$ \cite{wallman2015estimating, feng2016estimating}. Here, $u$ is the unitarity obtained by fitting an exponential model $\bar{P}_\mathrm{norm}(N_\mathrm{Cl}) = A'u^{N_\mathrm{Cl}} + B'$ to the averaged normalized purity computed for each sequence as $P_\mathrm{norm} = 2\mathrm{tr}(\hat{\rho}^2) - 1$ using the estimated density operator $\hat{\rho}$. As illustrated in Fig.~\ref{fig: stability and comparison}(b), the total error from RB, the incoherent error from purity RB and the leakage per gate remain practically constant over the studied time period, demonstrating that the FAST DRAG pulses enable the implementation of stable gates in our setup. As shown in the histograms of Fig~\ref{fig: stability and comparison}(c), the mean values of the total gate error, incoherent error and leakage per gate are $(1.60 \pm 0.01) \times 10^{-4}$, $(1.14 \pm 0.01) \times 10^{-4}$, and $(2.58 \pm 0.03) \times 10^{-5}$ suggesting that other sources contribute to the total gate error in addition to incoherent errors and leakage. We attribute the remaining errors to non-Markovian coherent errors resulting from microwave pulse distortions as discussed in more detail in the following subsection~\ref{sec: exp MW tails}.

\subsection{\label{sec: exp MW tails} Characterization and pre-compensation of control pulse distortions}

\begin{figure*}
\centering
\includegraphics[width = 0.92\textwidth]{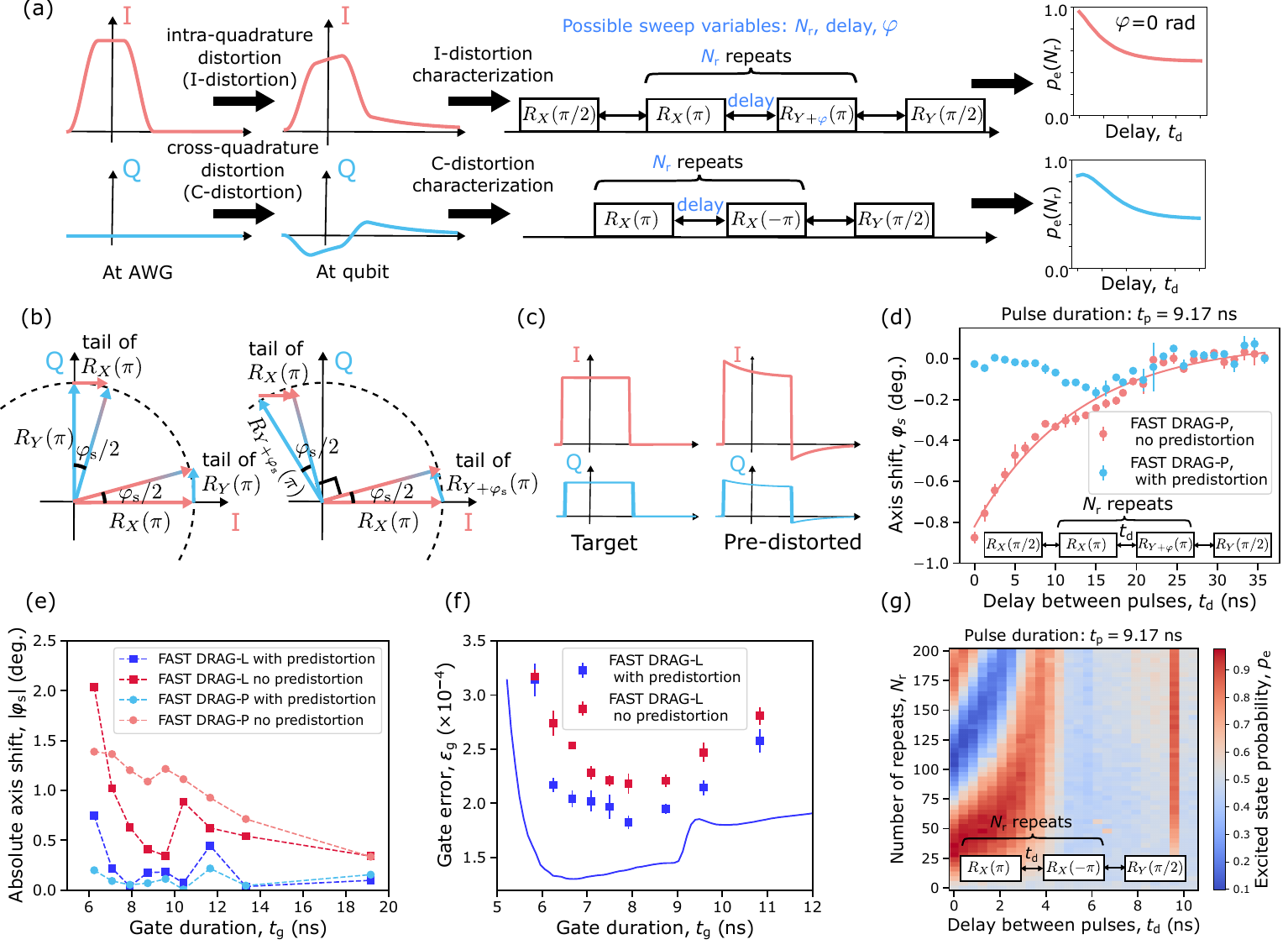}
    \caption{ \label{fig: MW distortions} \justifying \textbf{Characterization and mitigation of microwave pulse temporal distortions.} (a) Microwave pulse distortions are classified into intra-quadrature distortions (I-distortion, red) and cross-quadrature distortions (C-distortion, blue). 
    We experimentally characterize I- and C-distortions using different error-amplifying circuits with the measured excited state probability $p_\mathrm{e}$ being mainly sensitive to one type of distortion and depending on the delay $t_\mathrm{d}$ between consecutive pulses. 
    (b) Effective rotation axes (long arrows) of consecutive $R_X(\pi)$ (red) and $R_{Y + \varphi}(\pi)$ (blue) gates in I-distortion characterization when envelope tails (short arrows) due to I-distortion are present. The axis tilt $\varphi_\mathrm{s}$ caused by the envelope tails (left) can be cancelled by shifting the phase of the $R_{Y + \varphi}(\pi)$-gate with $\varphi = \varphi_\mathrm{s}$ (right). 
    (c) Schematic of envelope predistortion using an exponential IIR inverse filter to mitigate I-distortions for the $I$- (red) and $Q$-envelopes (blue). (d) Axis shift $\varphi_\mathrm{s}$ measured from I-distortion characterization as a function of the delay $t_\mathrm{d}$  between consecutive gates for a 9.17-ns-long FAST DRAG-P pulse with (light blue) and without (light red) predistortion. The error bars represent 1$\sigma$ uncertainty, and   
    the solid line shows an exponential fit to the data without predistortion. (e) Absolute axis shift $|\varphi_\mathrm{s}|$ as a function of gate duration for FAST DRAG-L (dark color) and FAST DRAG-P (light color) with (blue) and without (red) envelope predistortion when using the default  delay of $t_\mathrm{d}=0.41$ ns between pulses. 
    (f) Error per gate $\varepsilon_\mathrm{g}$ from RB (markers) and simulated error of $R_X(\pi/2)$ (solid line) as functions of the gate duration for FAST DRAG-L with (blue) and without (red) predistortion. 
    (g) Excited state probability as a function of delay $t_\mathrm{d}$  between consecutive pulses and the number of repeated $(R_X(\pi), R_X(-\pi))$-sequences in C-distortion characterization using FAST DRAG-P with envelope predistortion and $t_\mathrm{p}=9.17$~ns.  }
\end{figure*}

As mentioned in the previous section, coherent errors related to microwave pulse distortions appear to prevent us from implementing coherence-limited sub-10-ns gates. Here, we discuss experimental evidence of such pulse distortions using the qubit as a sensor and present our approach for predistorting control pulse envelopes to improve gate fidelity. We classify the pulse distortions into two types based on their effect on the pulse envelopes similarly to Ref.~\cite{gustavsson2013improving}, see Fig.~\ref{fig: MW distortions}(a). The first type of distortion, named intra-quadrature distortion (I-distortion), corresponds to a distortion of the in-phase (quadrature) envelope resulting from a pulse in the in-phase (quadrature) envelope. The second type of distortion, named cross-quadrature distortion (C-distortion), refers to a distortion of the quadrature (in-phase) envelope caused by a pulse in the in-phase (quadrature) envelope. Both types of distortions result in coherent errors, the effects of which depend on previous gates and the delay $t_\mathrm{d}$  between pulses in a way that cannot be corrected by optimizing the sample amplitudes during the gate, e.g., as in Ref. \cite{werninghaus2021leakage}. As such errors are non-Markovian, the gates cannot be described by fixed Pauli transfer matrices extracted, e.g., from gate set tomography \cite{nielsen2020probing, nielsen2021gate}, whereas the characterization of the full process tensor \cite{white2020demonstration, white2022characterization, white2022non} capturing non-Markovian errors would be time consuming. 
Here, we instead characterize the I- and C-distortions using straightforward error amplification experiments which are almost exclusively sensitive to only one type of distortion. 

To characterize C-distortions, we use a circuit consisting of repeated ($R_X(\pi), R_X(-\pi)$)-pairs as suggested in Ref.~\cite{gustavsson2013improving} and append the sequence with a final $R_Y(\pi/2)$-gate to make the sequence linearly sensitive to phase errors, see Fig.~\ref{fig: MW distortions}(a). As a result, the sequence is efficient in detecting C-distortions causing phase errors, the magnitude of which varies depending on the delay $t_\mathrm{d}$  between pulses. For the characterization of I-distortions, we propose a circuit consisting of repeated ($R_X(\pi), R_{Y+\varphi}(\pi)$)-pairs prepended with a $R_X(\pi/2)$ gate and appended with a $R_Y(\pi/2)$ gate to make the measured excited state probability linearly sensitive to distortion-induced coherent errors. In this experiment, each pair of  ($R_X(\pi), R_{Y}(\pi)$)-gates would ideally result in a $Z$-rotation.  
In the presence of I-distortions, the tail of the previous pulse tilts the rotation axis of the following pulse, resulting in a change of the relative angle between the axes of $R_X(\pi)$ and $R_Y(\pi)$, as shown in Fig~\ref{fig: MW distortions}(b). This further causes the resulting $Z$-rotation to have an over-rotation that can be amplified with repeated gates. Other common types of errors, such as amplitude miscalibration or phase errors, do not result in such overrotation to first order, but instead change the resulting rotation axis and thus do not contribute to the error amplification. We sweep the relative phase between $R_X(\pi)$ and $R_{Y+\varphi}(\pi)$ through $\varphi$ in order to determine an axis shift angle $\varphi_\mathrm{s}$ leading to a vanishing overrotation angle of the composite $Z$-rotation,  thus indicating cancellation of the axis tilt caused by the pulse tails as schematically illustrated in Fig~\ref{fig: MW distortions}(b). 
Therefore, we regard the axis shift $\varphi_\mathrm{s}$ as a measure of the magnitude of the I-distortions. Furthermore, we accompany the I- and C-distortion characterization experiments with a sweep of the delay between pulses to characterize the temporal profile of the pulse distortions.

We mitigate I-distortions by separately predistorting the $I$ and $Q$ control pulse envelopes using an exponential infinite impulse response (IIR) inverse filter in analogy to the predistortion of flux pulses \cite{foxen2018high, rol2020time,sung2021realization}, see Fig.~\ref{fig: MW distortions}(c) and Appendix~\ref{ap: predistortion} for more details. This choice of the filter function is motivated by the observation that the axis shift $\varphi_\mathrm{s}$ measured from the I-distortion characterization decays approximately exponentially as a function of the delay between pulses, as shown in Fig.~\ref{fig: MW distortions}(d). Using calibration measurements explained in Appendix~\ref{ap: predistortion}, we obtain a time constant $\tau_1 = 8$ ns and amplitude coefficient $a_1=-0.028$ as the predistortion parameters used for both $I$ and $Q$ envelopes. When using predistortion, we observe that the exponential dependence is removed in the I-distortion characterization resulting in a significantly smaller axis shift as shown in Fig.~\ref{fig: MW distortions}(d), providing support for the used predistortion model. In Fig~\ref{fig: MW distortions}(e), we further show that the predistortion significantly reduces the axis shift $\varphi_\mathrm{s}$ for FAST DRAG-P and FAST DRAG-L pulses across gate durations $t_\mathrm{g} \in [6, 20]$ ns. For FAST DRAG-L, the remaining axis shift is higher since the virtual Z-rotations preceding each pulse effectively convert part of the non-corrected C-distortion into I-distortion. In Fig.~\ref{fig: MW distortions}(f), we compare the RB gate error obtained using FAST DRAG-L pulses with and without predistortion and demonstrate that the gate error is systematically reduced by $(2-6)\times 10^{-5}$ across the tested gate durations between 6 ns and 11 ns.

When applying the predistortion, we still find a small coherent error as discussed above and illustrated in Fig.~\ref{fig: stability and comparison}(c). We attribute this remaining coherent error to C-distortions that we were not able to correct in this work, see Appendix~\ref{ap: predistortion}. To verify this conclusion, we conduct a C-distortion characterization experiment for varying pulse-to-pulse delays using a FAST DRAG-P pulse with $t_\mathrm{p} = 9.17$ ns. As shown in Fig.~\ref{fig: MW distortions}(g), the gates indeed suffer from delay-dependent phase errors that we attribute to remaining C-distortions. Note that we do not use the measurement scheme proposed in Ref.~\cite{gustavsson2013improving} to obtain the finite impulse response filter corresponding to the C-distortions since the assumption of vanishing gate duration does not hold well for a transmon-based system unlike for the flux qubit studied in Ref.~\cite{gustavsson2013improving}.

\section{\label{sec: outlook} Outlook}

In this work, we propose and demonstrate two new methods for shaping the frequency spectrum of single-qubit control pulses, enabling stronger spectral suppression of leakage transitions and thus lower leakage errors for fast logic gates compared to state-of-the-art approaches. Using the proposed methods, we calibrate $R_X(\pi/2)$-gates and experimentally demonstrate a gate error of $(1.56 \pm 0.07) \times 10^{-4}$ at a gate duration of 7.9 ns without experimental closed-loop optimization, outperforming conventional approaches both in terms of gate error and speed. 

The developed pulse shaping methods are not only limited to superconducting qubits or to the suppression of the $ef$-transition of a single transmon qubit, but they may also be applied to other platforms or to suppress multiple leakage transitions or other undesired transitions. Thus, the proposed methods may improve simultaneous single-qubit gates on quantum processing units of any type that suffer from frequency crowding issues. This may be achieved by minimizing the spectral energy of the control pulses across crosstalk-induced transitions in addition to the $ef$-transition, while using a cutoff frequency to regularize the pulse shape. Even though HD DRAG and FAST DRAG provide an approximately equal leakage error in the studied single-qubit system, we foresee that the flexibility of FAST DRAG may be beneficial to appropriately adjust the relative strength and width of spectral suppression in the case of multiple undesired transitions as discussed in Appendix~\ref{ap: hyperparameter selection}.

The proposed methods may also find applications in the optimization of two-qubit gate control pulses
based on microwave pulses \cite{rigetti2010fully, Ding2023_FTF} or flux pulses \cite{sung2021realization,marxer2023long}. As an example, architectures based on the cross-resonance effect \cite{kim2023evidence, malekakhlagh2020first, li2024experimental} or the recently demonstrated fluxonium-transmon-fluxonium scheme \cite{Ding2023_FTF} rely on microwave-activated two-qubit gates that suffer from frequency crowding. Hence, the analytical pulse shaping methods developed in this work may help to suppress undesired transitions improving the microwave-activated two-qubit gates, and potentially speeding up the calibration compared to, e.g., the reinforcement learning approach employed in Ref.~\cite{Ding2023_FTF}. 
Furthermore, we expect that combining the proposed methods with experimental closed-loop optimization may result in further improvements to the fidelity of fast quantum logic gates in systems with high coherence. Namely, the proposed methods provide good initial parameter guesses and a natural basis for the optimization, potentially reducing the required number of iterations and wall clock time significantly. 

We also demonstrate that non-Markovian coherent errors arising from microwave pulse distortions contribute to the gate error of sub-10-ns single-qubit gates. Hence, we develop and demonstrate methods to observe errors caused by such distortions  and subsequently apply control pulse predistortion to reduce gate errors. However, our approach did not mitigate the pulse distortions completely, and further work is needed to develop methods for more precise characterization of such distortions using the qubit as a sensor to enable improved predistortion of the control pulses. All in all, we expect that our work contributes to improving the fidelity and speed of quantum logic gates, 
bringing the advent of useful quantum computation one step closer. 

\section*{Data availability}
Data supporting the findings of this article is available at: \url{https://doi.org/10.5281/zenodo.13365033}. The dataset in Zenodo also includes a Jupyter notebook that provides Python helper functions for evaluating and plotting the FAST DRAG and HD DRAG pulse shapes defined by Eqs.~\eqref{eq: FAST cos series}, \eqref{eq: c from matrix eq}, \eqref{eq: 3rd der DRAG I}, and \eqref{eq: 3rd der DRAG Q}. 

\begin{acknowledgments}

We thank the whole staff at IQM Quantum Computers for their support. Especially, we acknowledge Lucas Ortega, Roope Kokkoniemi and Matthew Sarsby for supporting the maintenance and improvement of the experimental setup; Tuure Orell, Jani Tuorila, and Hao Hsu for simulation-related discussions; and Attila Geresdi on discussions related to single-qubit gates.
E.H. thanks the Finnish Foundation for Technology Promotion (grant No. 9230) and Jenny and Antti Wihuri Foundation (grant No. 00230115) for funding. We acknowledge the provision of facilities and technical support by Aalto University at OtaNano - Micronova Nanofabrication Center and LTL infrastructure.

All authors declare that IQM has filed a patent application regarding the new methods for shaping the frequency spectrum of control pulses having the following inventors: Eric Hyyppä, Antti Vepsäläinen and Johannes Heinsoo.

E.H. developed the concept and theory for FAST DRAG and HD DRAG. E.H., A.V., and J.H. planned the simulations and the experiments. E.H. conducted the simulations with support from M.P. and S.I.. E.H. conducted the experiments and analyzed the measurement data with support from J.L., F.M., S.O., and C.F.C. on measurement code,  F.M. on experiment setup, and B.T. on experiment software. A.L. and C.O.-K designed the sample. W.L. fabricated the device. The manuscript was written by E.H. with support from A.V., and J.H.. All authors commented on the manuscript. A.V., and J.H. supervised the work.  

\end{acknowledgments}

\appendix

\section{\label{ap: FAST ap} Derivation of FAST pulse shapes}

Here, we provide a more detailed derivation for Eq.~\eqref{eq: c from matrix eq} that allows us to solve the coefficients $\{c_n \}$ of a FAST pulse from a given set of weights and frequency intervals to be suppressed. At the end of this section, we also briefly discuss some useful properties of the pulse shaping method in more detail compared to the main text.

In the FAST method, the control pulse envelope $\Omega_\mathrm{I}(t)$ with a duration $t_\mathrm{p}$ is expressed as a finite sum of basis functions 
\begin{align}
\Omega_\mathrm{I}(t) &= A \times \sum_{n=1}^N c_n g_n(t) \label{seq: I FAST DRAG in terms of g},
\end{align}
where $N$ is the number of basis functions, $\{g_n(t)\}$ is the set of basis functions defined to be 0 outside of the interval $[0, t_\mathrm{p}]$, and $\{c_n\}$ are the corresponding coefficients. In this work, we have chosen the sum of basis functions to represent a Fourier cosine series parametrized such that the control envelope $\Omega_\mathrm{I}:  \mathbb{R} \rightarrow \mathbb{R}$ and its derivative have no discontinuities, thus reducing slow sinc-type decay of the pulse spectrum. Thus, we write the control envelope $\Omega_\mathrm{I}(t)$ as
\begin{align}
\Omega_\mathrm{I}(t) = A \times \sum_{n=1}^N c_n [ 1 - \cos(2\pi n t / t_\mathrm{p}) ] \Pi(t/t_\mathrm{p} - 1/2), \label{seq: I FAST DRAG in terms of cos} 
\end{align}
where $\Pi(x)$ is the rectangular window function between $x\in[-1/2, 1/2]$, which sets the pulse envelope to 0 outside of $[0, t_\mathrm{p}]$.

The coefficients $\{c_n\}$ are obtained by minimizing the spectral energy of the pulse across undesired frequency intervals corresponding to, e.g., leakage transitions. We solve for the coefficients $\{c_n\}$ by employing the following linearly-constrained quadratic optimization problem
\begin{gather}
    \textrm{minimize}~\sum_{j=1}^k w_j \int_{f_{\mathrm{l}, j}}^{f_{\mathrm{h}, j}} |\hat{\Omega}_\mathrm{I}(f)|^2 \mathrm{d}f \label{seq: FAST min cost function}, \\
    \mathrm{s.t.}~\sum_{n=1}^N c_n \times t_\mathrm{p} = \theta,
\end{gather}
where $w_j$ represents the weight, i.e., importance, for the $j$th frequency interval $[f_{\mathrm{l}, j}, f_{\mathrm{h}, j}]$, $k$ is the number of frequency interval to suppress, and $\theta$ is the desired rotation angle that can be chosen arbitrarily since the amplitude parameter $A$ is at any rate adjusted in experimental calibration. Note that the frequencies are in the baseband, i.e., measured with respect to the center drive frequency, and  the spectrum of the modulated microwave control pulse will consequently be minimized across frequency intervals $f_\mathrm{d} \pm [f_{\mathrm{l}, j}, f_{\mathrm{h}, j}]$ symmetrically located around the central drive frequency $f_\mathrm{d}$. 

By expanding the control envelope $\Omega_\mathrm{I}(t)$ using Eq.~\eqref{seq: I FAST DRAG in terms of g}, the optimization problem in Eq.~\eqref{seq: FAST min cost function} can be written in a matrix form as
\begin{align}
    &\sum_{j=1}^k w_j \int_{f_{\mathrm{l}, j}}^{f_{\mathrm{h}, j}} |\hat{\Omega}_I(f)|^2 \mathrm{d}f \nonumber \\
    &= \sum_{j=1}^k w_j \int_{f_{\mathrm{l}, j}}^{f_{\mathrm{h}, j}} \left(\sum_{n=1}^N c_n \hat{g}_n(f) \right) \left(\sum_{m=1}^N c_m \hat{g}_m^*(f) \right) \mathrm{d}f \nonumber \\
    &= \sum_{n,m} c_n c_m \left(\sum_{j=1}^k w_j \int_{f_{\mathrm{l}, j}}^{f_{\mathrm{h}, j}} \hat{g}_n(f) \hat{g}_m^*(f)  \mathrm{d}f \right) \nonumber \\
    &= \sum_{n,m} c_n c_m \bm{A}_{nm} \nonumber \\
    &= \bm{c}^T \bm{A} \bm{c}, \label{seq: weighted sum as mat format}
\end{align}
where $\bm{c}\in \mathbb{R}^{N \times 1}$ is a column vector containing the coefficients $c_n$ as its elements, and $\bm{A} \in \mathbb{C}^{N \times N}$ is a hermitian matrix with elements $\bm{A}_{nm} = \sum_{j=1}^k w_j \int_{f_{\mathrm{l}, j}}^{f_{\mathrm{h}, j}} \hat{g}_n(f) \hat{g}_m^*(f)  \mathrm{d}f$. Here, the Fourier transform $\hat{g}_n(f)$ of the $n$th basis function can be analytically solved as
\begin{align}
    \hat{g}_n(f) &= \int_{-\infty}^{\infty} g_n(t) \mathrm{e}^{-\mathrm{i} 2\pi tf} \mathrm{d} t \\
    &= \int_{0}^{t_\mathrm{p}} [1 - \cos(2\pi n t/t_\mathrm{p})] \mathrm{e}^{-\mathrm{i} 2\pi tf} \mathrm{d} t \nonumber \\
    &= t_\mathrm{p} \bigg(\mathrm{e}^{-\mathrm{i} \pi t_\mathrm{p} f} \mathrm{sinc}(\pi t_\mathrm{p} f) \nonumber \\
    &- \frac{1}{2}\mathrm{e}^{\mathrm{i} \pi (n/t_\mathrm{p}- \nonumber f)t_\mathrm{p}}\mathrm{sinc}[\pi(n/t_\mathrm{p} - f) t_\mathrm{p}] \nonumber \\
    &-  \frac{1}{2}\mathrm{e}^{-\mathrm{i} \pi (n/t_\mathrm{p} + f)t_\mathrm{p}}\mathrm{sinc}[\pi(n/t_\mathrm{p} + f) t_\mathrm{p}] \bigg),
\end{align}
where $\mathrm{sinc}(x) = \sin(x)/x$. Importantly, the analytic form of $\hat{g}_n(f)$ ensures efficient computation of the matrix elements $\bm{A}_{nm}$ via numerical integration. Using Eq.~\eqref{seq: weighted sum as mat format}, the optimization problem can then be re-written in a matrix form as
\begin{gather}
    \textrm{minimize}~\bm{c}^T \bm{A} \bm{c}\\
    \mathrm{s.t.}~ \bm{c}^T\bm{b} - \theta/t_\mathrm{p} = 0,
\end{gather}
where we have further defined $\bm{b} = (1, \dots, 1)^T \in \mathbb{R}^{N \times 1}$. 

Now, the minimization problem can be analytically solved using the method of Lagrangian multipliers. The Lagrangian corresponding to the optimization problem reads $L(\bm{c}, \mu) = \bm{c}^T \bm{A} \bm{c} - \mu(\bm{c}^T\bm{b} - \theta/t_\mathrm{p} )$, and the optimality conditions are thus given by 
\begin{gather}
    \nabla_{\bm{c}} L = (\bm{A} + \bm{A}^T)\bm{c} - \mu \bm{b} = 0 \\
    \nabla_{\mu} L = \bm{c}^T\bm{b} - \theta/t_\mathrm{p} = 0.
\end{gather}
These two conditions can be combined into a single matrix equation
\begin{equation}
    \tilde{\bm{A}} \tilde{\bm{c}} = \tilde{\bm{b}}, \label{seq: FAST matrix equation}
\end{equation}
where $\tilde{\bm{c}} = (\bm{c}^T, \mu)^T$ is the coefficient vector extended by the Lagrangian multiplier, $\tilde{\bm{b}} = (\bm{0}^T, \theta/t_\mathrm{p})^T \in \mathbb{R}^{(N+1)\times 1}$ with $\bm{0} = (0, \dots, 0)^T \in \mathbb{R}^{N \times 1}$, and the matrix $\tilde{\bm{A}} \in \mathbb{R}^{(N+1)\times (N+1)}$ is defined as
\begin{equation}
    \tilde{\bm{A}} = \begin{pmatrix}
    \bm{A} + \bm{A}^T & -\bm{b} \\
    \bm{b}^T & 0
    \end{pmatrix}.
\end{equation}
This concludes the derivation of Eq.~\eqref{eq: c from matrix eq}. 

Subsequently, we briefly discuss some useful properties of the FAST method. See also Appendix~\ref{ap: hyperparameter selection} for guidelines on the selection of the hyperparameters. 
First, we emphasize that the FAST method enables the construction of pulses with a given duration. Thus, the method differs from filter-based solutions, such as the use of band-pass or low-pass filters \cite{mckay2017efficient}, which may filter out specific frequency intervals from the spectrum of a control pulse but which also result in pulse distortions, potentially significantly changing the pulse duration. As already discussed in Sec.~\ref{sec: FAST}, the FAST method enables one to control the width and strength of the spectral suppression. Thus, a combination of FAST and DRAG may be helpful to reach a lower leakage error for fast logic gates compared to conventional DRAG pulses since the spectral suppression provided by conventional DRAG pulses may not be sufficient, e.g., in the presence of multiple leakage transitions or frequency shifts occurring during the gate. However, one should bear in mind that the suppression of several frequency intervals using the FAST method may lead to increased spectral power at non-suppressed frequencies.

As another important property of the FAST method, it is possible to limit the bandwidth of the pulse by suppressing the spectrum across an additional frequency interval ranging from a given cutoff frequency $f_\mathrm{c}$ up to some high frequency. This not only helps to abide by the bandwidth limitations of the control electronics, but may also help to reduce the peak amplitude of the pulse as observed in Fig. ~\ref{fig: leakage vs duration res}(f) avoiding compression of analog electronics and heatload to the cryostat. Reducing the cutoff frequency effectively favors pulses with a lower average power. In the limit of suppressing the spectrum across $[f_{\mathrm{l}, 1}, f_{\mathrm{h}, 1}]=[0, \infty]$, one obtains the minimum-energy pulse for realizing a rotation with the given angle in the given time, essentially approximating a square pulse with the given number of harmonic terms. 

As a further benefit of the FAST method, the most important parameters $(A, \beta, \{[f_{\mathrm{l}, j}, f_{\mathrm{h}, j}]\})$ of a single-qubit control pulse  $\Omega_\mathrm{RF}(t) = \Omega_\mathrm{I}(t)\cos(\omega_\mathrm{d}t) -\beta \dot{\Omega}_I(t)/\alpha \sin(\omega_\mathrm{d}t)$ are practically independent. This means that the rotation angle of the gate is mainly controlled by the amplitude parameter $A$, whereas $\beta$ and $\{[f_{\mathrm{l}, j}, f_{\mathrm{h}, j}]\}$ independently control which parts of the frequency spectrum are suppressed since the Fourier transform of the complex envelope $\Omega_\mathrm{IQ}$ is given by the product form $\hat{\Omega}_\mathrm{IQ}(f)=[1 - 2\pi\beta f/\alpha]\times \hat{\Omega}_I(f)$. The  baseband frequency intervals $\{[f_{\mathrm{l}, j}, f_{\mathrm{h}, j}]\}$ are symmetrically suppressed around the center drive frequency, and have thus no effect on the resulting effective drive frequency (unlike the DRAG parameter $\beta$). 

Finally, we note that other choices for the basis functions $\{g_n(t)\}$ beyond the cosine series are also possible and the coefficients $\{c_n\}$ of the basis functions can still be solved using Eq.~\eqref{seq: FAST matrix equation}. The expression for $\hat{g}_n(f)$ needs to be analytically evaluated for the new set of basis functions or numerically computed across the undesired frequency intervals, which may, however, increase the computational cost of solving $\{c_n\}$.

\section{\label{ap: HD DRAG ap} Higher-derivative DRAG}

In this section, we provide a more general formulation of the HD DRAG framework beyond the special case containing derivatives up to the third order as discussed in Sec.~\ref{sec: higher-derivative DRAG}. Using the first $2K+1$ derivatives for the in-phase and quadrature envelopes, the envelope functions can be written as
\begin{align}
        \Omega_\mathrm{I}(t) &= A \bigg[\sum_{n=0}^K \beta_{2n} g^{(2n)} (t) \bigg], \label{eq: HD DRAG I} \\
    \Omega_\mathrm{Q}(t) &= -\frac{\beta }{\alpha} \dot{\Omega}_\mathrm{I} , \label{eq: HD DRAG Q}
\end{align}
where $g(t)$ is a basis envelope, and $\{\beta_{2n}\}_{n=0}^{2K}$ are coefficients for the even-order derivatives with $\beta_0=1$. To avoid leakage transitions or other harmful frequencies, we require the Fourier transform $\hat{\Omega}_\mathrm{I}(f)$ to equal zero at $K$ different baseband frequencies $\{f_j\}_{j=1}^K$
\begin{equation}
    \hat{\Omega}_\mathrm{I}(f_j) = 0,~j=1, \dots, K. \label{seq: HD omegaI zero}
\end{equation}
Using the relation $\mathcal{F}[s^{(n)}](f) = (\imag 2\pi f)^n \hat{s}(f)$ with $\mathcal{F}[\cdot]$ denoting Fourier transform, we simplify the expression of the Fourier transform  $\hat{\Omega}_\mathrm{I}(f)$  as
\begin{equation}
    \hat{\Omega}_\mathrm{I}(f) = A\hat{g}(f) \bigg [\sum_{n=0}^K \beta_{2n} (-1)^n (2\pi f)^{2n} \bigg  ]. \label{eq: Omega I hat polynomial}
\end{equation}
Thus, the condition in Eq.~\eqref{seq: HD omegaI zero} reduces to the following system of linear equations
\begin{equation}
    \sum_{n=0}^K \beta_{2n} (-1)^n (2\pi f_j)^{2n} = 0,~j=1, \dots, K,  \label{eq: equation of betas for HD DRAG}
\end{equation}
from which the coefficients $\{\beta_{2n}\}_{n=1}^{K}$ can be straightforwardly solved assuming that the suppressed frequencies $\{f_j\}_{j=1}^K$ are distinct. However, it is also possible to suppress the spectrum  more heavily around a single baseband frequency, such as $\alpha/(2\pi)$, for $K > 1$ by engineering the polynomial in Eq.~\eqref{eq: Omega I hat polynomial} to have a $K$th order zero at $\alpha/(2\pi)$. For $K=2$ with $\hat{\Omega}_\mathrm{I}$ being a fourth-order polynomial in $f$, this can be achieved by setting $\beta_2 = 2/\alpha^2$ and $\beta_4 = 1/\alpha^4$.

The formulation of HD DRAG in Eqs.~\eqref{eq: HD DRAG I} and \eqref{eq: HD DRAG Q} decouples the in-phase parameters $\{\beta_{2n}\}_{n=1}^{K}$ and associated suppressed frequencies $\{f_j\}_{j=1}^K$ from the DRAG coefficient $\beta$, which simplifies the calibration experiments especially for the case of $K=1$ considered in Sec.~\ref{sec: higher-derivative DRAG} and also enables one to tune $\beta$ to avoid phase errors if desired (see Sec.~\ref{sec: DRAG}). By solving both $\{\beta_{2n}\}_{n=1}^{K}$ and $\beta$ to suppress a given undesired baseband frequency $f_\mathrm{s}$, a stronger and wider spectral suppression can be obtained than with conventional DRAG pulses.  Furthermore, the in-phase parameters $\{\beta_{2n}\}_{n=1}^{K}$ do not affect the effective drive frequency (unlike $\beta$), which can be beneficial for calibration experiments in some scenarios. 

As discussed in Sec.~\ref{sec: higher-derivative DRAG}, the basis function $g(t)$ should be chosen with care to ensure continuity and avoid slow decay of the Fourier transform as a function of frequency. One possibility is to use a cosine series
\begin{equation}
    g(t) = \sum_{k=1}^{K + 1} d_k[1 - \cos(2\pi k t/t_\mathrm{p})], \label{seq: g cosine series} 
\end{equation}
where $t_\mathrm{p}$ is the pulse duration and $\{d_k\}_{k=1}^{K+1}$ denote the coefficients to be solved, such that the continuity requirements $g(0)=g^{(1)}(0)=\dots=g^{(2K+1)}(0)=g(t_\mathrm{p})=g^{(1)}(t_\mathrm{p})=\dots=g^{(2K+1)}(t_\mathrm{p})=0$ are satisfied. Due to the form of $g(t)$ in Eq. ~\eqref{seq: g cosine series} the odd derivatives are automatically zero, whereas the conditions for the even derivatives can be met by solving $\{d_k\}_{k=1}^{K+1}$ from the following linear system of equations 
\begin{align}
    \sum_{k=1}^{K+1} d_k (-1)^{n+1} k^{2n} &= 0, ~n=1, \dots, K, \\
    \sum_{k=1}^{K+1} d_k &= 1,
\end{align}
where the latter equation simply normalizes the sum of the coefficients to a value that can be set arbitrarily. Even though this approach ensures the smoothness of the control pulse and was found to enable fast, low-leakage single-qubit gates with up to $2K+1=3$ derivatives, the higher-order solutions may lead to rapidly oscillating envelope functions, the bandwidth and peak amplitude of which cannot be regularized unlike with the proposed FAST method.

\section{\label{ap: 1qb gate simulations} Single-qubit gate simulations}

In this section, we describe the methods used to simulate the gate error and leakage as illustrated in Fig.~\ref{fig: leakage vs duration res}, and in Supplementary Figs.~\ref{sfig: leakage for all pulses}$-$\ref{sfig: 3rd-der DRAG sweeps}. For the simulations, we model the bare transmon using an anharmonic Duffing oscillator Hamiltonian as 
\begin{equation}
    \Ham_\mathrm{q}/\hbar = \omegaq \ad \ah + \frac{\alpha}{2} \ad \ad \ah \ah, \label{eq: bare qubit}
\end{equation}
where $\omegaq/(2\pi)\approx 4.417$ GHz and $\alpha/(2\pi) \approx -212$ MHz are the qubit frequency and anharmonicity, and $\ad$ and $\ah$ are the raising and lowering operators of a harmonic oscillator. The capacitively coupled microwave drive is modeled by the following drive term 
\begin{equation}
    \Ham_\mathrm{d}/\hbar = \Omega_\mathrm{RF}(t) \left( \ad + \ah \right), \label{eq: drive hamiltonian}
\end{equation}
where the drive signal is of the form $\Omega_\mathrm{RF}(t) = \Omega_\mathrm{I}(t) \cos(\omegad t + \varphi) +  \Omega_\mathrm{Q}(t) \sin(\omegad t + \varphi) $ with $\omegad/(2\pi)$ being the drive frequency and $\varphi$ describing the accumulated phase due to Virtual Z-rotations. 

For the simulations, we transform to a frame rotating at the qubit frequency using the unitary $U=\exp(\imag \omegaq \ad \ah t)$ and make the rotating wave approximation resulting in the following Hamiltonian 
\begin{align}
     \Ham_\mathrm{R}/\hbar &=  \frac{\alpha}{2} \ad \ad \ah \ah + \frac{1}{2} \bigg \{ \ad \mathrm{e}^{-\imag(\omegad - \omegaq)t}[\tilde{\Omega}_I(\varphi, t) + \imag \tilde{\Omega}_Q(\varphi, t)] \nonumber \\
     &+ \ah \mathrm{e}^{\imag(\omegad - \omegaq)t} [\tilde{\Omega}_I(\varphi, t) - \imag \tilde{\Omega}_Q(\varphi, t)] \bigg \}, \label{seq: Ham sim RWA} 
\end{align}
where $\tilde{\Omega}_I(\varphi, t)= \Omega_\mathrm{I}(t)\cos\varphi +  \Omega_\mathrm{Q}(t)\sin\varphi$ and $\tilde{\Omega}_Q(\varphi, t)= -\Omega_\mathrm{I}(t)\sin\varphi +  \Omega_\mathrm{Q}(t)\cos\varphi$ incorporate the accumulated phase $\varphi$. In addition to the coherent dynamics governed by the Hamiltonian in Eq.~\eqref{seq: Ham sim RWA}, our simulation incorporates decoherence by solving the system dynamics from the following Lindblad master equation using the \verb|mesolve|-function in QuTiP \cite{johansson2012qutip}
\begin{equation}
    \dot{\rho} = -\frac{\imag}{\hbar}[\Ham_\mathrm{R}, \rho] + \sum_i \bigg [L_i\rho L_i^\dagger - \frac{1}{2}\left(\rho L_i^\dagger L_i +  L_i^\dagger L_i \rho \right) \bigg], \label{seq: Lindbladian}
\end{equation}
where $\rho$ denotes the density operator, and we use the jump operators $L_{-}=\sqrt{1 + \bar{n}}/\sqrt{T_1}\ah$, $L_{+}=\sqrt{\bar{n}}/\sqrt{T_1}\ad$, and $L_\varphi = 1/\sqrt{T_\varphi}\ad \ah$ to model thermal relaxation, thermal excitation, and dephasing, respectively \cite{wiseman2009quantum,wood2018quantification}. For the simulations, we set the thermal steady-state population as $\bar{n}=0.02$  based on single-shot readout experiments, the relaxation time as $T_1=35$ $\mu$s, and the dephasing time as $T_\varphi=40$ $\mu$s in line with experimental results. In the simulations, we further truncate the state space of the anharmonic oscillator to the four lowest-energy levels. We use four levels instead of three since we observe that the three-level simulation sometimes underestimates the leakage error in comparison to experimental results for short gate durations. 

To simulate the application of a single-qubit $R_X(\pi/2)$-gate for given control pulse envelopes $\Omega_\mathrm{I}$ and $\Omega_\mathrm{Q}$, we evolve the system dynamics according to Eq.~\eqref{seq: Lindbladian} starting from a known pure initial state, and include a zero padding of 0.41 ns after each pulse as in the experiments. Importantly, we aim to study the leakage and fidelity under ideal pulses and thus we do not model non-idealities of the control electronics, such as pulse distortions or a finite sampling rate, dynamic range or bandwidth. In the case of control pulses based on DRAG-L, we advance the phase $\varphi$ by $\varphi_z/2$ before and after each pulse to correct for phase errors using virtual Z-rotations \cite{mckay2017efficient}. As a result of the applied virtual Z-rotations during $N_\mathrm{gates}$ gates, a final $Z$-rotation by $-\varphi=-N_\mathrm{gates}\varphi_z$ emerges \cite{mckay2017efficient}, which does not affect experimentally measured probabilities since the measurement occurs along the $Z$-axis. However, this extra $Z$-rotation needs to be undone for the simulated density operator using an additional $Z$-rotation by $\varphi$ to evaluate the gate fidelity using the density operator. 

For a given control pulse shape and gate duration, we set the drive frequency equal to the qubit frequency and calibrate the pulse parameters $\{A, \beta, \varphi_z\}$ for a $R_X(\pi/2)$-gate based on DRAG-P or DRAG-L using a simulated calibration sequence similar to the one used in the experiments and described in Appendix~\ref{ap: 1qb gate calib}. As a minor difference, the leakage per gate needed for calibrating $\beta$ in the case of DRAG-L is estimated based on a single simulated $R_X(\pi/2)$-gate instead of using a leakage RB sequence.

After the calibration,  the system dynamics are simulated for several initial states, and the error per gate is approximated as 
\begin{equation} 
    \varepsilon_\mathrm{g} \approx 1-  \frac{1}{6} \sum_i \langle \psi_i | R_X(\pi/2)^\dagger \rho(\psi_i) R_X(\pi/2)|\psi_i \rangle,
\end{equation}
where $\rho(\psi_i)$ is the resulting density operator (including the final $Z$-correction)  starting from the initial state $\psi_i$, and the sum of the initial states goes over the six cardinal states $\{|0\rangle, |1\rangle, (|0\rangle + |1\rangle)/\sqrt{2}, (|0\rangle - |1\rangle)/\sqrt{2}, (|0\rangle + \imag|1\rangle)/\sqrt{2}, (|0\rangle - \imag|1\rangle)/\sqrt{2} \}$. The leakage per gate is subsequently computed as the average probability to leave the computational subspace across the six cardinal initial states, i.e.,
\begin{equation} 
    L \approx \frac{1}{6} \sum_i \{ 1 - \mathrm{Tr}[|0\rangle \langle 0| \rho(\psi_i)] - \mathrm{Tr}[|1\rangle \langle 1| \rho(\psi_i)]  \}.
\end{equation}

\section{\label{ap: exp setup} Experimental setup}

\begin{figure}
\centering
\includegraphics[width = 0.48\textwidth]{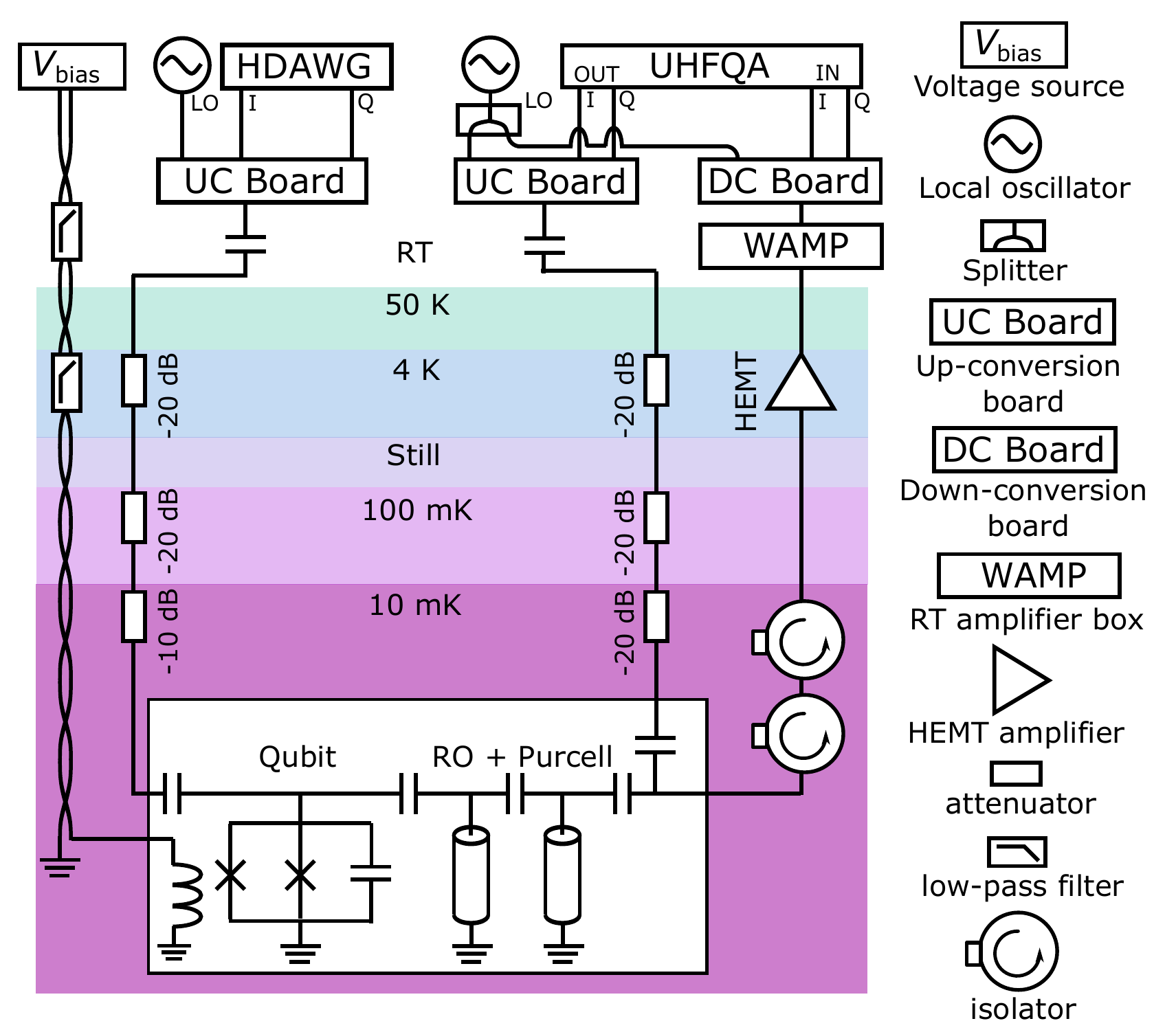}
\caption{ \label{sfig: exp setup} \justifying \textbf{Wiring diagram of the experimental setup.} Schematic of the experimental setup used in the single-qubit gate experiments showing one of the two flux-tunable qubits on the used test device. Brief explanations for some of the symbols are presented in the column on the right. 
}
\end{figure}

The experimental setup used for the single-qubit gate experiments is illustrated in the wiring diagram of Fig.~\ref{sfig: exp setup}. For the experiments, a two-qubit test device is cooled down to 10-mK base temperature using a commercial dilution refrigerator. The used test device has similar design as that used in Ref.~\cite{marxer2023long}, in which more details are provided regarding target design parameters and sample fabrication. Both of the qubits on the device are grounded flux-tunable transmons, the frequencies of which can be controlled through individual on-chip flux lines connected to a room-temperature voltage source via a twisted-pair cable and a pair of low-pass filters. 

To implement single-qubit rotations, the qubits are capacitively coupled to individual drive lines that are connected to room-temperature electronics via 50 dB of total nominal attenuation separated across different temperature stages of the cryostat. At the qubit frequency, the total room-temperature attenuation in the cryostat was characterized to be approximately 67 dB. At room temperature, the single-qubit control pulses are generated at an intermediate frequency using a channel pair of a HDAWG (Zurich Instruments) with a sampling rate of 2.4 GSa/s and a peak-to-peak maximum amplitude range of 3 V or 4 V enabled by an internal amplifier, which limits the analog bandwidth to 300 MHz. To attain a sample-level granularity for the gate duration and facilitate envelope predistortion discussed in Appendix~\ref{ap: predistortion}, we evaluate the $I$ and $Q$ envelope waveforms of each gate sequence on a computer accounting for the Virtual Z-rotations, after which we send these long waveforms to the HDAWG.   The intermediate-frequency control pulses are upconverted to the qubit frequency using an IQ mixer and an in-house-built local oscillator with an integrated root-mean-squared phase jitter below 4 mrad from 1 kHz to 10 MHz. 

For qubit state readout, the qubits are capacitively coupled to individual readout structures consisting of a readout resonator and a Purcell filter with both of the readout structures further coupled to a common probe transmission line. We generate the readout pulses at an intermediate frequency using an UHFQA (Zurich Instruments)  and upconvert the pulses with an IQ mixer and a single SGS100A microwave generator (Rohde \& Schwarz) that is used both for the up-conversion and down-conversion of the readout signal. On its way to the sample, the readout signal is passed through 60 dB of nominal attenuation. The readout signal transmitted through the probe line is amplified using an high-electron-mobility transistor (HEMT) at the 4 K stage with further amplification applied at room temperature. The amplified readout signal is downconverted back to an intermediate frequency and subsequently digitized using the UHFQA. Note that our setup does not include a quantum-limited parametric amplifier despite which we reach a decent two-state readout assignment fidelity of up to $\sim 95$\%.

\section{\label{ap: 3state ro} Three-state readout}

\begin{figure}[h]
\centering
\includegraphics[width = 0.37\textwidth]{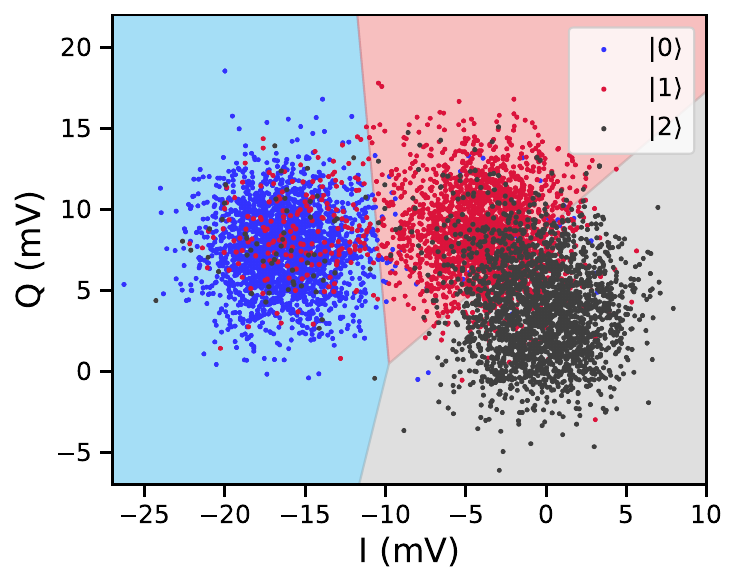}
\caption{ \label{sfig: 3state ro} \justifying \textbf{Three-state classification.} Single-shot readout voltages in the IQ plane for the prepared states $|0\rangle$ (blue dots), $|1\rangle$ (red dots) and $|2\rangle$ (black dots). The colored regions illustrate the state classification boundaries.
To compute the readout assignment matrix, 
we use $2^{15}=32768$ shots for each of the prepared states, but we have down-sampled the number of points by a factor of 15 for illustration purposes.}
\end{figure}

In this section, we briefly describe our approach for performing three-state readout on the qubit used for the single-qubit gate studies. At the flux-insensitive operation point, we measure a dispersive shift of $\chi/(2\pi) = 4.0$ MHz. For the readout pulse, we use a readout frequency of 6.146 GHz and a relatively long pulse duration of 1.5~$\mu$s due to the absence of a quantum-limited amplifier in the readout chain. For the integration of the acquired readout signal, we use an integration time of 1.5 $\mu$s and set all integration weights to unity. 
To calibrate the classification boundaries for three-state readout, we repeatedly prepare the qubit in each of the states $|0\rangle$, $|1\rangle$, and $|2\rangle$, and record $2^{15} = 32678$ single-shot IQ voltages for each state, as shown in Fig.~\ref{sfig: 3state ro}. For each prepared state, the mean of the acquired IQ voltages is computed and for new readout events, the qubit is assigned to the state with the smallest distance between the recorded single-shot IQ voltage and the state mean.  The resulting readout assignment matrix is given by 
\begin{equation}
    \bm{\beta} = \begin{bmatrix} 0.972 & 0.025 & 0.003 \\
    0.095 & 0.742 & 0.162 \\
    0.024 & 0.126 & 0.850
    \end{bmatrix},
\end{equation}
where  $\bm{\beta}_{ij}$ corresponds to the probability of assigning the qubit to the $j$th state after preparing it to the $i$th state. 
We attribute the readout errors to finite state overlap between the states $|1\rangle$ and $|2\rangle$, state preparation errors due to a thermal $e$-state population of $\sim 2$\%, and $T_1$ decay during the readout.

Similarly to, e.g., Ref.~\cite{mckay2017efficient}, we correct most of the remaining readout errors  using the inverse of the state assignment matrix as
\begin{equation}
    \bm{p}_\mathrm{corr} = \bm{\beta}^{-T} \bm{p}_\mathrm{meas},
\end{equation}
where $\bm{p}_\mathrm{corr}=(p_g, p_e, p_f) \in \mathbb{R}^{3\times 1}$ is the corrected vector of state probabilities, and $\bm{p}_\mathrm{meas}  \in \mathbb{R}^{3\times 1}$ is the measured vector of state probabilities before the correction. We point out that the readout correction is important for accurate measurement of leakage per gate 
since the leakage RB protocol is sensitive to SPAM errors unlike the standard RB and interleaved RB protocols. 

\section{\label{ap: hyperparameter selection} Hyperparameter selection for FAST DRAG and HD DRAG}

In this section, we provide guidelines for selecting and calibrating the hyperparameters for FAST DRAG ($N$, $\{w_j\}$, $\{[f_{\mathrm{l}, j}, f_{\mathrm{h}, j}]\}$) and HD DRAG ($\{ \beta_{2n} \}$). We first consider a scenario where the $ef$-transition is the dominant source of leakage as in this work. Subsequently, we briefly cover the expected modifications that are needed to mitigate multiple leakage transitions, e.g., in simultaneous single-qubit gates suffering from crosstalk.

\begin{figure}
\centering
\includegraphics[width = 0.4\textwidth]{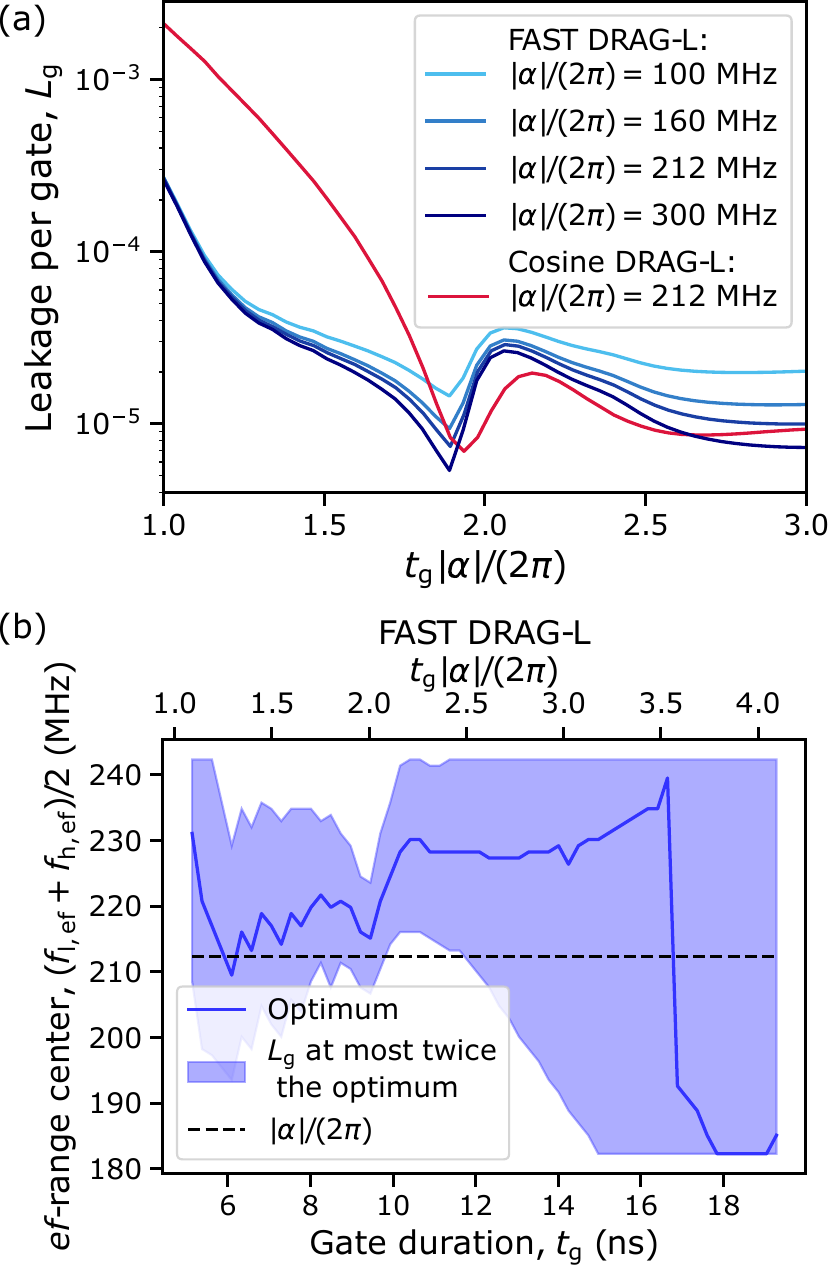}
    \caption{ \label{sfig: hyperparam sweep} \justifying \textbf{Hyperparameter selection for FAST DRAG.} (a) Simulated leakage per gate $L_\mathrm{g}$ as a function of the normalized gate duration $t_\mathrm{g}|\alpha|/(2\pi)$ for different anharmonicities $\alpha/(2\pi) \in \{-100, -160, -212, -300\}$ MHz (shades of blue) when selecting the hyperparameters of FAST DRAG-L using the heuristic algorithm of Appendix~\ref{ap: hyperparameter selection}. For comparison, we show the leakage of Cosine DRAG-L (red line) for $\alpha/(2\pi) = -212$ MHz. In the simulations, we assume $T_1 = 35$~$\mu$s, $T_\varphi = 40$~$\mu$s, and $\bar{n} = 0.02$ as explained in Appendix~\ref{ap: 1qb gate simulations}. (b) Optimal value for the center of the suppressed frequency interval $(f_{\mathrm{l}, \mathrm{ef}} + f_{\mathrm{h}, \mathrm{ef}}) / 2$ (dark blue) to minimize simulated leakage per gate $L_\mathrm{g}$ as a function of the gate duration for FAST DRAG-L when selecting other parameters based on the heuristic algorithm of Appendix~\ref{ap: hyperparameter selection}. The colored area shows values of $(f_{\mathrm{l}, \mathrm{ef}} + f_{\mathrm{h}, \mathrm{ef}}) / 2$ yielding a leakage per gate at most twice the optimal value, while the dashed line corresponds to $|\alpha|/(2\pi)$. For each gate duration, the DRAG coefficient $\beta$ is calibrated assuming $(f_{\mathrm{l}, \mathrm{ef}} + f_{\mathrm{h}, \mathrm{ef}}) / 2=|\alpha|/(2\pi)$ and kept fixed while sweeping $(f_{\mathrm{l}, \mathrm{ef}} + f_{\mathrm{h}, \mathrm{ef}}) / 2$.  }
\end{figure}

To mitigate leakage to the $f$-level, the hyperparameters of the FAST method can be selected using the following physics-based heuristic algorithm:
\begin{enumerate}
    \item Parametrize the weight ratio as $w_\mathrm{ef} = w_1/w_2$ and the frequency intervals to be suppressed as $[f_{\mathrm{l}, 1}, f_{\mathrm{h}, 1}]=[f_{\mathrm{l}, \mathrm{ef}}, f_{\mathrm{h}, \mathrm{ef}}]$ and $[f_{\mathrm{l}, 2}, f_{\mathrm{h}, 2}]=[f_{\mathrm{c}}, \infty]$. Now, the hyperparameters to be determined are $\{f_{\mathrm{l}, \mathrm{ef}}, f_{\mathrm{h}, \mathrm{ef}}, f_{\mathrm{c}}, w_\mathrm{ef}, N\}$.
    \item Measure the anharmonicity  $\alpha/(2\pi)$ of the qubit in Hz. 
    %\item Set the center of the first frequency interval to the measured anharmonicity, and the width, e.g. as 20 MHz, i.e., $f_{\mathrm{l}, \mathrm{ef}} = |\alpha|/(2\pi) - 10~\mathrm{MHz}$ and $f_{\mathrm{h}, \mathrm{ef}} = |\alpha|/(2\pi) + 10~\mathrm{MHz}$.  
    \item Set the center of the first frequency interval to the measured anharmonicity and the width, e.g., as $0.1 \times |\alpha|/(2\pi)$ ($\sim 20$ MHz for typical transmon anharmonicities) resulting in $f_{\mathrm{l}, \mathrm{ef}} = 0.95 \times |\alpha|/(2\pi)$ and $f_{\mathrm{h}, \mathrm{ef}} = 1.05 \times |\alpha|/(2\pi)$. %For typical transmon anharmonicities, an interval width of $0.1 \times |\alpha|/(2\pi) \sim 20$~MHz works well according to our experiments and simulations.
    \item Set the cutoff frequency as $f_\mathrm{c} = 2|\alpha|/(2\pi)$ based on the measured anharmonicity.
    \item Set the weight as $w_\mathrm{ef} = 5$ for DRAG-L and $w_\mathrm{ef} = 100$ for DRAG-P similarly to the experiments in this work. 
    \item The number of Fourier terms can be set, e.g., as $N=4$.
    \item To implement single-qubit gates, use Eqs.~\eqref{eq: FAST cos series} and \eqref{eq: c from matrix eq} to obtain the pulse shape for the selected hyperparameters
\end{enumerate}
For HD DRAG, the procedure is simpler as the only additional parameter is $\beta_2$, which can be set as $\beta_2 = 1/\alpha^2$ to mitigate $f$-level leakage. According to simulations shown in Fig.~\ref{sfig: hyperparam sweep}(a), the above guidelines are expected to outperform Cosine DRAG-L at least for typical transmon anharmonicities $|\alpha|/(2\pi) \in (150, 300)$ MHz across fast gate durations $t_\mathrm{g}|\alpha|/(2\pi) \in (1, 2)$, for which Cosine DRAG-L is not sufficient to suppress leakage errors. Note that the hyperparameters obtained with these guidelines are very close to the parameters used for the experiments in Sec.~\ref{sec: exp results}.

The above heuristic guidelines are based on the following reasoning. In the experiments and the simulations, the optimal center frequency $(f_{\mathrm{l}, \mathrm{ef}} + f_{\mathrm{h}, \mathrm{ef}}) / 2$ is observed to approximately equal the anharmonicity $|\alpha|/(2\pi)$, with the minimum of leakage being relatively broad, as shown in Figs.~\ref{sfig: FAST DRAG-P sweeps}(e) and \ref{sfig: FAST DRAG-L sweeps}(e) in Appendix~\ref{ap: FAST param sweeps}. By selecting $(f_{\mathrm{l}, \mathrm{ef}} + f_{\mathrm{h}, \mathrm{ef}}) / 2 = |\alpha|/(2\pi)$, one is expected to obtain a leakage rate that is typically within a factor of two of the optimal leakage rate, see Fig.~\ref{sfig: hyperparam sweep}(b). 
The choice of the cutoff frequency $f_\mathrm{c} = 2|\alpha|/(2\pi)$ provides a reasonable balance between leakage suppression and regularization of the pulse shape, which ensures a low leakage error for fast gates while limiting the peak amplitude compared to Cosine DRAG-L pulses (see Fig.~\ref{fig: leakage vs duration res}(f)). When it comes to the weight $w_\mathrm{ef}$ and the interval width $f_{\mathrm{h}, \mathrm{ef}} - f_{\mathrm{l}, \mathrm{ef}}$, we observe that the resulting leakage error for FAST DRAG-L pulses is largely insensitive to the exact value of these parameters both in simulations and experiments as long as the parameters have the correct order of magnitude, see Figs.~\ref{sfig: FAST DRAG-L sweeps}(d) and (f) in Appendix~\ref{ap: FAST param sweeps}. However, we observe that too aggressive spectral suppression with an excessively high value of $w_\mathrm{ef}$ may lead to increased leakage errors with the DRAG-L calibration, as shown in Fig.~\ref{sfig: FAST DRAG-L sweeps}(d). 

\begin{figure}
\centering
\includegraphics[width = 0.42\textwidth]{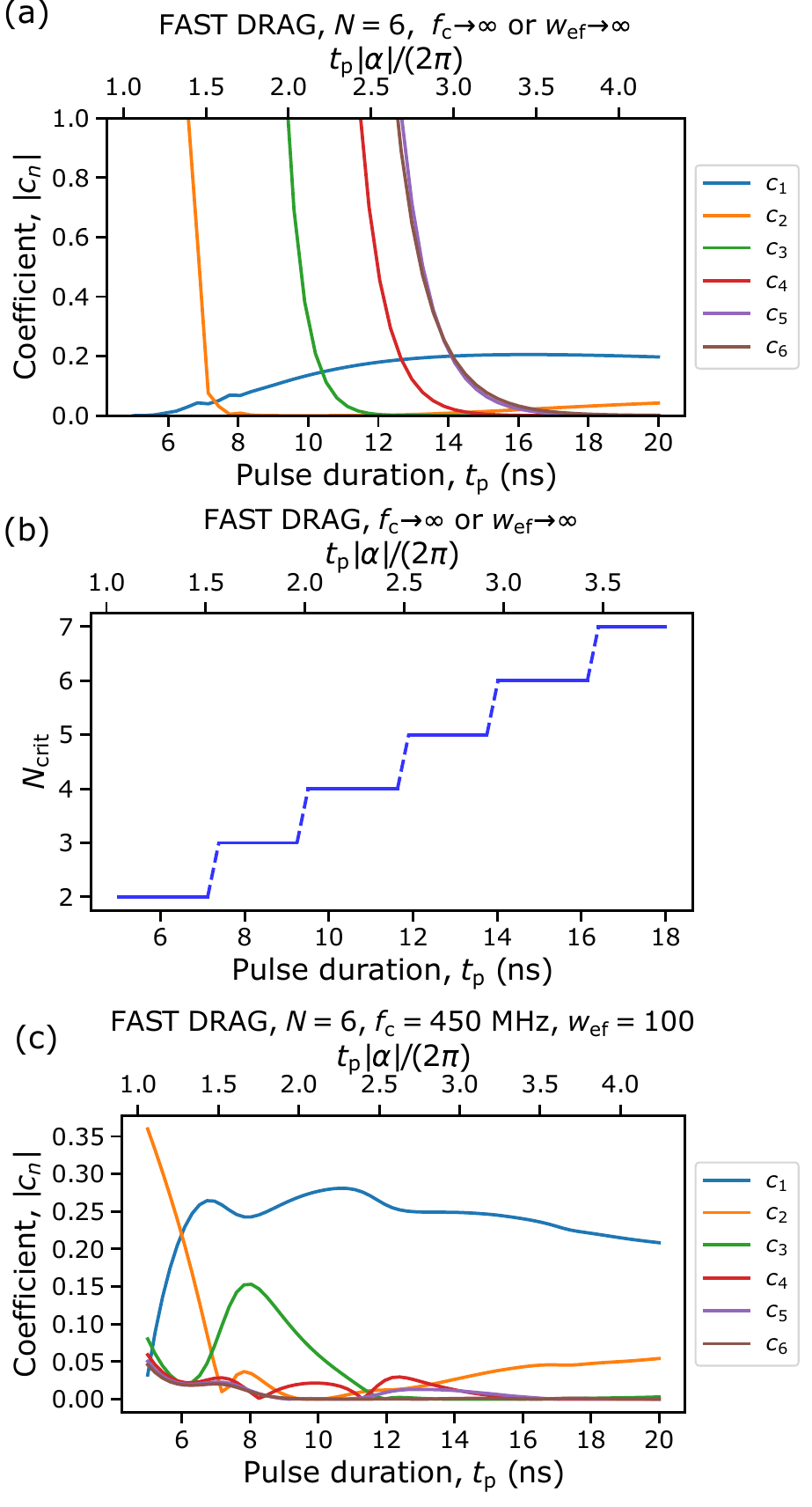}
    \caption{ \label{sfig: Fourier coefs vs tg} \justifying \textbf{Regularizing effect of the cutoff frequency on the Fourier coefficients in FAST DRAG pulses.} (a) Absolute value of the Fourier coefficients $|c_n|$ as a function of pulse duration for a FAST DRAG pulse with $N=6$ and a single suppressed frequency interval $[f_{\mathrm{ef}, \mathrm{l}}, f_{\mathrm{ef}, \mathrm{h}}] = [194, 214]$ MHz, i.e.,  $f_\mathrm{c} \to \infty$ or $w_\mathrm{ef} \to \infty$. (b) Critical number of basis functions $N_\mathrm{crit}$ as a function of pulse duration for a FAST DRAG pulse with $f_\mathrm{c} \to \infty$ or $w_\mathrm{ef} \to \infty$. (c) Same as (a) but using a second frequency interval to specify a cutoff frequency $f_\mathrm{c}=450$ MHz and a weight $w_\mathrm{ef} = 100$ controlling the relative suppression across the two frequency intervals. }
\end{figure}

Furthermore, there is some freedom in the selection of the number of Fourier terms since already $N=2$ provides a low leakage rate for FAST DRAG-L in the studied system with one dominant leakage transition, as shown in Fig.~\ref{sfig: FAST DRAG-L sweeps}(b). However, a higher number of basis functions $N \geq 3$ may be beneficial to suppress the in-phase spectrum across more than one leakage transitions or to control the strength of the spectral suppression. This is exemplified in Fig.~\ref{sfig: FAST DRAG-P sweeps}(b) demonstrating that the leakage errors caused by FAST DRAG-P pulses are reduced by increasing the number of basis functions, which allows a stronger spectral suppression around the $ef$-transition for fast gates. 
Importantly,  the use of a finite cutoff frequency $f_\mathrm{c}$ and weight $w_\mathrm{ef}$ render the pulse shape and the leakage error less sensitive to the choice of $N$. For $f_\mathrm{c} \to \infty$ or $w_\mathrm{ef} \to \infty$, there exists a critical value $N_\mathrm{crit}$ below which the choice of $N$ naturally limits the bandwidth and peak power but above which the Fourier coefficients $\{c_n\}$ attain high absolute values leading to a high peak power and bandwidth of the control pulse.  Here, we define the critical value $N_\mathrm{crit}$ as the largest value of $N$ ensuring that the highest-frequency component remains bounded according to the condition $|c_N| < \theta/t_\mathrm{p}$  in the case of $f_\mathrm{c} \to \infty$ or $w_\mathrm{ef} \to \infty$, where $\theta$ is the rotation angle of Eq.~\eqref{eq: FAST constraint}
As the pulse duration is reduced, the Fourier coefficients of high-frequency components increase first as illustrated in Fig.~\ref{sfig: Fourier coefs vs tg}(a), thus implying that the critical value $N_\mathrm{crit}$ is reduced for short gate durations, as shown in Fig.~\ref{sfig: Fourier coefs vs tg}(b). For fast gates with $t_\mathrm{p}|\alpha|/(2\pi) \in (1, 2)$, the number of basis functions would need to be limited to $N=2$ to ensure a reasonable bandwidth and peak power in the case of $f_\mathrm{c} \to \infty$ or $w_\mathrm{ef} \to \infty$. For these reasons, we favor the use of a finite cutoff frequency and weight, which lead to bounded Fourier coefficients across gate durations of interest, see Fig.~\ref{sfig: Fourier coefs vs tg}(c). Thus, the cutoff frequency allows us to regularize the pulse shape, while still benefiting from the stronger and more flexible spectral suppression and potentially reduced leakage errors obtained with increasing $N$. 

We note that further performance improvements compared to the heuristic algorithm may be achieved by performing closed-loop optimization of the hyperparameters as a part of the single-qubit gate calibration. The parameter values predicted by the heuristic algorithm can be used as educated initial values to enhance convergence.

In the presence of multiple leakage transitions, e.g., in simultaneous single-qubit gates, the heuristic algorithm for FAST DRAG needs to be extended. Similarly to the single-qubit case, one should measure the frequencies of potential undesired transitions, and estimate the coupling strength to each of these transitions. Subsequently, one can choose to suppress frequency intervals corresponding to the $ef$-transition and the dominant crosstalk-induced transitions, while reserving an additional frequency interval to regularize the pulse shape. A good starting point for FAST DRAG is to set the center frequencies of the intervals equal to the frequencies of the undesired transitions. To enable strong spectral suppression across more than one frequency intervals, one should also use $N\geq 3$ basis functions. Furthermore, one should be aware that aggressive spectral suppression across multiple frequency intervals leads to increased spectral power across non-suppressed frequencies and a wider bandwidth, which may have unintended consequences in systems with high crosstalk or with many transitions to avoid.  Thus, further research is needed to design reliable heuristics for setting the weights and widths of the frequency intervals using the measured frequencies and the associated coupling strengths. With HD DRAG, the suppressed baseband frequencies $\{f_j\}$ can be directly set to the undesired transition frequencies, after which the parameters $\{\beta_{2n}\}$ can be solved from Eq.~\eqref{eq: equation of betas for HD DRAG}. As emphasized in Sec.~\ref{sec: higher-derivative DRAG}, HD DRAG cannot control the relative strength of spectral suppression between the different frequency intervals in contrast to FAST DRAG that provides more flexibility, which is expected to enable an enhanced performance in systems with multiple leakage transitions. As the multi-qubit system is more challenging and the required number of parameters is higher compared to the single-qubit system,  closed-loop optimization may be needed to reach close-to-optimal performance. However, physics-based heuristics are expected to help in finding good initial values for the optimization.

\section{\label{ap: 1qb gate calib} Single-qubit gate calibration}

\begin{figure}
\centering
\includegraphics[width = 0.5\textwidth]{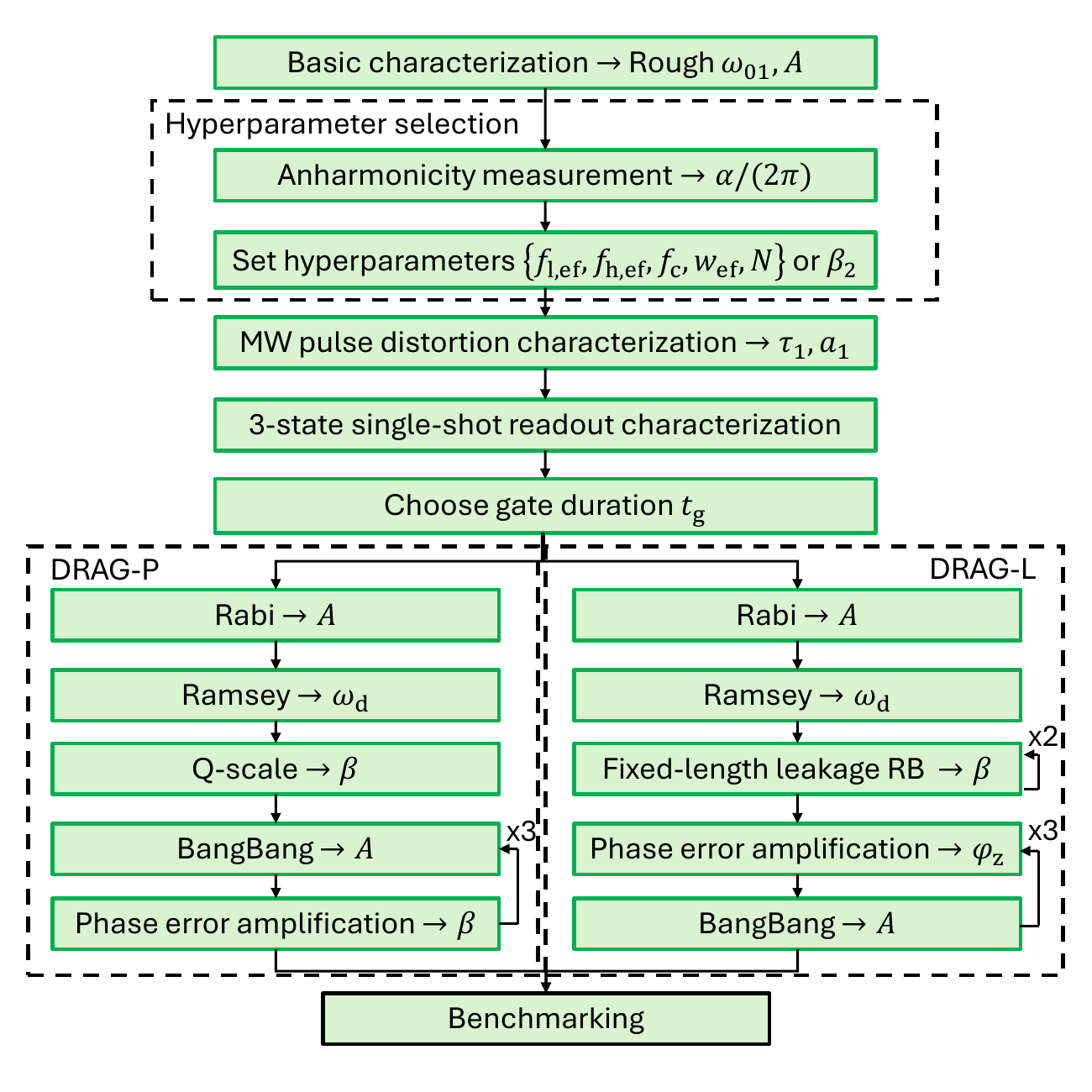}
    \caption{ \label{sfig: calibration flow} \justifying \textbf{Calibration flow chart.} Flow chart of the single-qubit gate calibration. See Appendix~\ref{ap: predistortion} and Fig.~\ref{sfig: pre-dist calib} for more details on microwave pulse predistortion calibration and Appendix~\ref{ap: hyperparameter selection} for more details on hyperparameter selection. }
\end{figure}

Here, we describe our procedure for calibrating the parameters of a single-qubit $R_X(\pi/2)$-gate implemented with a control pulse of the form $\Omega_\mathrm{RF}(t) = \Omega_\mathrm{I}(t) \cos(\omegad t + \varphi) +  \Omega_\mathrm{Q}(t) \sin(\omegad t + \varphi) $, where $\Omega_\mathrm{I}(t) = A \times \Omega_\mathrm{I}^\mathrm{norm}(t)$ with $A$ being the in-phase amplitude,  $\Omega_\mathrm{Q}(t) = -\beta\dot{\Omega}_I(t)/\alpha$, and the phase $\varphi$ is either set to zero for DRAG-P or advanced by $\varphi_z$ before each gate for DRAG-L. Different calibration experiments are needed for DRAG-P and DRAG-L as DRAG-P tunes $\beta$ to minimize phase errors, whereas DRAG-L calibrates $\beta$ to minimize leakage as discussed in Sec.~\ref{sec: DRAG}. See Fig.~\ref{sfig: calibration flow} for a flow chart of the calibration procedure.

For DRAG-P, we calibrate the set of parameters $\{A, \beta, \omega_\mathrm{d}/(2\pi)\}$, such that the amplitude $A$ controls the rotation angle of the gate, $\beta$ minimizes phase errors during the gate and $\omega_\mathrm{d}/(2\pi)$ is set to the qubit frequency to avoid phase errors during idling. Assuming the qubit frequency is roughly known, we first determine an initial estimate for the amplitude $A$ by sweeping its value in a Rabi oscillation experiment. Subsequently, the qubit frequency $\omegaq/(2\pi)$ is obtained from a Ramsey experiment, and the drive frequency is set as $\omega_\mathrm{d}/(2\pi)=\omegaq/(2\pi)$. An initial estimate for the DRAG coefficient $\beta$ is then obtained using a Q-scale experiment, in which the excited state probability is recorded for gate sequences $(R_X(\pi), R_Y(\pi/2))$ and $(R_Y(\pi), R_X(\pi/2))$ for varying $\beta$ to find the value of $\beta$ with equal excited state probability after both sequences, which ideally corresponds to vanishing phase errors \cite{reed2013entanglement}. Finally, the estimates for $A$ and $\beta$ are improved by iterating a loop of error amplification measurements. In the loop, the amplitude estimate $A$ is refined in a BangBang experiment, in which the excited state probability is measured after a gate sequence consisting of an initial $R_X(\pi/2)$-gate followed by repeated ($R_X(\pi/2)$, $R_X(\pi/2)$, $R_X(\pi/2)$, $R_X(\pi/2)$)-sequences amplifying amplitude errors. By sweeping the amplitude $A$ and the number of rotations around the Bloch sphere, the optimal value of $A$ is accurately determined based on the measured excited state probability. Subsequently, we conduct a phase error amplification experiment consisting of repeated ($R_X(\pi/2)$, $R_X(\pi/2)$, $R_X(-\pi/2)$, $R_X(-\pi/2)$)-sequences \cite{chen2016measuring, bengtsson2020quantum} followed by a final $R_Y(\pi/2)$-gate, thanks to which the measured excited state probability is linearly sensitive to phase errors \cite{werninghaus2022experimental}. By sweeping $\beta$ and the number of repeated sequences, the value of $\beta$ minimizing phase errors can be accurately measured. Typically, we iterate the loop of error amplification experiments 2-3 times, and the whole calibration procedure lasts a few minutes with significant room for improvement using, e.g., restless measurements \cite{rol2017restless, haupt2023restless} or active qubit reset.

For DRAG-L, we calibrate the set of parameters $\{A, \beta, \varphi_z, \omega_\mathrm{d}/(2\pi)\}$ such that the amplitude $A$ controls the rotation angle of the gate, $\beta$ minimizes leakage, $\varphi_z$ counteracts the accumulated phase error during the gate, and $\omega_\mathrm{d}/(2\pi)$ is set to equal the qubit frequency to avoid phase errors during idling. Similarly to the case of DRAG-P, we first determine an initial estimate for $A$ from a Rabi oscillation experiment followed by a characterization of the qubit frequency from a Ramsey experiment. Subsequently, a leakage RB experiment is conducted with a fixed number of Clifford gates and the $f$-state population after the RB sequence is measured as a function of $\beta$ to minimize the $f$-state leakage. The leakage RB experiment with a sweep of $\beta$ is  repeated  with a longer Clifford sequence and a narrower sweep range to improve the estimate of $\beta$. Finally, the virtual-Z parameter $\varphi_z$ and the amplitude $A$ are estimated from an iterative loop of error amplification experiments: First, the value of $\varphi_z$ minimizing phase errors is determined by sweeping $\varphi_z$ and the number of repeated gates in a phase error amplification experiment. Subsequently, the amplitude $A$ is obtained from a similar BangBang experiment as used for DRAG-P. Typically, we iterate the loop 2-3 times, and the whole calibration procedure lasts approximately 10 minutes with the leakage calibration dominating the calibration time. For both DRAG variants, the commonly employed AllXY sequence \cite{reed2013entanglement} may be used as a simple check whether the calibration has been successful.

\begin{figure*}[t]
\centering
    \includegraphics[width = 0.92\textwidth]{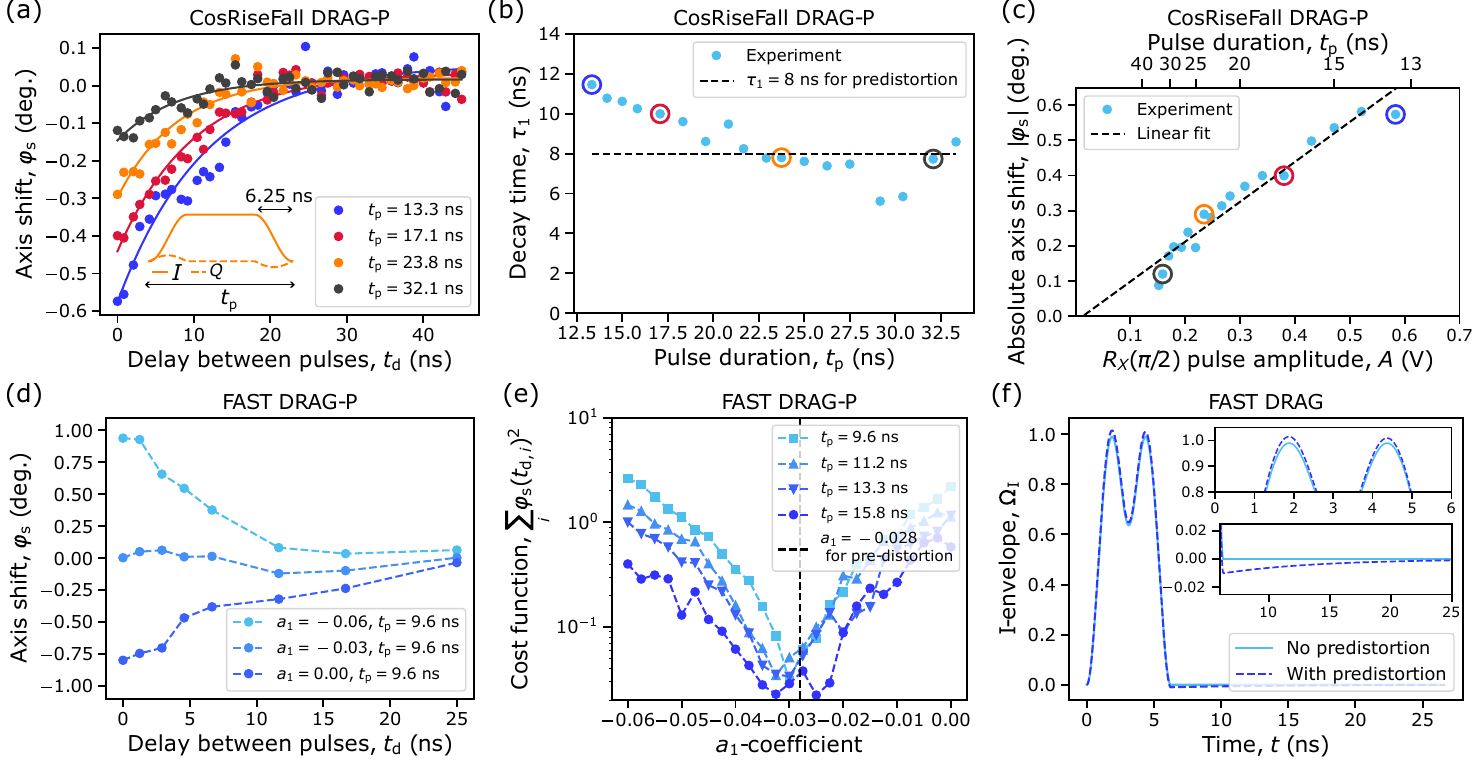}
    \caption{ \label{sfig: pre-dist calib} \justifying \textbf{Parameter calibration for envelope predistortion using an exponential IIR inverse filter.}
    (a) Axis shift $\varphi_\mathrm{s}$ (circles) measured from I-distortion characterization (see  Fig.~\ref{fig: MW distortions}(a)) as a function of the delay $t_\mathrm{d}$ between consecutive pulses for pulse durations $t_\mathrm{p} \in \{13.3, 17.1, 23.8, 32.1\}$ ns. The solid lines show exponential fits estimating the time constant $\tau_1$ for each pulse duration. The inset shows the used pulse shape, the in-phase envelope of which has a 6.25-ns-long cosine-shaped rise and fall. 
    (b) Decay time constant $\tau_1$ measured as in panel (a) as a function of the pulse duration $t_\mathrm{p}$. The blue, red, orange and grey circles correspond to the results for the pulse durations of panel (a). The dashed line shows $\tau_1=8$ ns chosen for exponential IIR predistortion. (c) Absolute axis shift $|\varphi_\mathrm{s}|$ without any delay between consecutive pulses, i.e., $t_\mathrm{d}=0$ as a function of pulse duration and thus pulse amplitude. The dashed line shows a linear fit to data. (d) Axis shift $\varphi_\mathrm{s}$ from I-distortion characterization  as a function of delay $t_\mathrm{d}$ between consecutive pulses for three values of the predistortion coefficient $a_1$ when using a FAST DRAG-P pulse with $t_\mathrm{p} = 9.6$ ns. 
    (e) Cost function $\sum_i \varphi_\mathrm{s} (t_{\mathrm{d}, i})^2$ as a function of  $a_1$ for four different pulse durations of FAST DRAG-P, with the minimum of the cost function at $a_1 = -0.028$ indicated by the dashed line. 
    (f) In-phase envelope for a 6.25-ns-long FAST DRAG pulse with (dashed dark blue) and without (solid light blue) envelope predistortion. The insets show close-ups to the peaks and tail of the pulse. 
    }
\end{figure*}

\section{\label{ap: predistortion} Implementation and calibration of control pulse predistortion.}

In this section, we provide more details on our approach for predistorting the control pulse envelopes. Similarly to Ref.~\cite{gustavsson2013improving}, we apply the framework of linear time-invariant (LTI) systems to model the distortions of the control pulse envelopes occurring between the generation of the control pulse and the arrival of the pulse to the qubit device. For performing the predistortion, we more specifically assume that both the $I$ and $Q$ control envelopes suffer from I-distortions (see Fig.~\ref{fig: MW distortions}) modeled with a continuous-time exponential IIR filter. We expect the step response at the qubit to be 
\begin{equation}
    y_i(t) = h(t) \ast u_i(t) = u_i(t) \times (1 + \sum_j a_j \mathrm{e}^{-t/\tau_j}),  
\end{equation}
where for the component $i\in \{\mathrm{I}, \mathrm{Q} \}$, $h(t)$ is the impulse response, i.e., the kernel function assumed equal for both components, $u_i(t)$ is a step-function, $\{a_j\}$ is the set of amplitude coefficients and $\{\tau_j\}$ are the respective decay time constants. We need to predistort the generated control pulse envelopes as $x_{i, \mathrm{pred}}(t) = h^{-1}(t) \ast x_{i, \mathrm{target}}(t)$ such that the control pulse envelopes reaching the qubit match the targeted envelope functions $x_{i, \mathrm{target}}(t)$. In practice, it is more convenient to implement the predistortion in the frequency domain \cite{gustavsson2013improving}, in which convolutions are converted into products and the Fourier transform of the predistorted control pulse envelope is given by 
\begin{equation}
    \hat{x}_{i, \mathrm{pred}}(f) =\frac{\hat{x}_{i, \mathrm{target}}(f)}{\hat{h}(f)}, \label{eq: FT of pred envelope}
\end{equation}
with the Fourier transform of the exponential IIR kernel function $\hat{h}(f)$ given by
\begin{equation}
    \hat{h}(f) = 1 + \sum_j \frac{\imag a_j 2\pi f \tau_j}{1 + \imag 2\pi f \tau_j }. \label{eq: FT of h}
\end{equation}
In our predistortion approach, we only use a single exponential with parameters $a_1$ and $\tau_1$.

In practice, the predistortion procedure is as follows \cite{gustavsson2013improving}: First, the targeted waveforms $x_{\mathrm{I}, \mathrm{target}}[n]$ and $x_{\mathrm{Q}, \mathrm{target}}[n]$ for the $I$ and $Q$ components  are constructed  based on the gate sequence to be executed, and sufficient zero padding is added to the end of the waveforms.  
Subsequently, the discrete Fourier transforms $\hat{x}_{\mathrm{I}, \mathrm{target}}[f]$ and $\hat{x}_{\mathrm{Q}, \mathrm{target}}[f]$ are computed using the Fast Fourier Transform (FFT) algorithm, and the Fourier transforms for the predistorted envelopes $\hat{x}_{\mathrm{I}, \mathrm{pred}}[f]$ and $\hat{x}_{\mathrm{Q}, \mathrm{pred}}[f]$ are obtained using Eqs. \eqref{eq: FT of pred envelope} and \eqref{eq: FT of h}. Finally, the predistorted waveforms $x_{\mathrm{I}, \mathrm{pred}}[n]$ and $x_{\mathrm{Q}, \mathrm{pred}}[n]$  are obtained in the time domain by taking the inverse discrete Fourier transform.

To calibrate the predistortion parameters $\{\tau_1, a_1\}$  and to validate the use of the exponential IIR filter model for I-distortions, we use the I-distortion characterization experiment consisting of repeated $(R_X(\pi), R_{Y+\varphi}(\pi)) = (R_X(\pi/2), R_X(\pi/2), R_{Y+\varphi}(\pi/2), R_{Y+\varphi}(\pi/2))$-sequences that amplify coherent errors caused by I-distortions, while being practically insensitive to other common coherent errors, such as amplitude miscalibration or phase errors as explained in Sec.~\ref{sec: exp MW tails}. First, we characterize the decay time constant $\tau_1$ by measuring the axis shift $\varphi_\mathrm{s}$ from an I-distortion characterization experiment as a function of the delay $t_\mathrm{d}$ between consecutive pulses without applying predistortion as illustrated in Fig.~\ref{sfig: pre-dist calib}(a). To approximate a square pulse and the resulting exponential tail, we parametrize the in-phase envelope as
\begin{align}
    \Omega_\mathrm{I}(t) = \begin{cases} 
    [ 1 - \cos(2\pi t/(2t_\mathrm{r}) ] / 2, ~&t \in [0, t_\mathrm{r}) \\
    1, ~&t \in [ t_\mathrm{r},  t_\mathrm{p} - t_\mathrm{r}) \\
    [ 1 - \cos(2\pi (t_\mathrm{p} - t)/(2t_\mathrm{r}) ] / 2,~ &t \in [t_\mathrm{p} - t_\mathrm{r}, t_\mathrm{p}], 
    \end{cases}
\end{align}
where $t_\mathrm{r}=6.25$ ns is a fixed rise time. The quadrature envelope is obtained using DRAG-P to avoid virtual Z-rotations used by DRAG-L from mixing I- and C-distortions. 

As shown in Fig.~\ref{sfig: pre-dist calib}(a), we estimate the decay time $\tau_1$ from an exponential fit to the results of an I-distortion characterization experiment, which we repeat for varying gate durations. 
As shown in Fig.~\ref{sfig: pre-dist calib}(b), the decay times 
vary between 6 ns and 12 ns with the results being less reliable for pulse durations above 30 ns due to a reduced signal-to-noise ratio (see, e.g., the grey points in Fig.~\ref{sfig: pre-dist calib}(a)). Based on the data, the measured decay time is approximately 8 ns across a wide range of pulse durations $t_\mathrm{p} \in [18, 28]$ ns, for which the square pulse approximation is reasonable while the signal-to-noise ratio is still high. Hence, we choose to use $\tau_1=8$ ns for the exponential IIR predistortion. In Fig.~\ref{sfig: pre-dist calib}(c), we further show the axis shift $|\varphi_\mathrm{s}|$ measured without delay between pulses as a function of the pulse duration and the corresponding pulse amplitude, which allows us to verify that the axis shift $|\varphi_\mathrm{s}|$ is linearly proportional to the pulse amplitude as expected for distortions caused by an LTI filter.  

Having determined the decay time $\tau_1$, we calibrate the amplitude coefficient $a_1$. To achieve this, we again measure the axis shift $\varphi_\mathrm{s}$ as a function of the delay $t_\mathrm{d}$ between pulses in an I-distortion characterization experiment but this time we predistort the control envelopes using the calibrated decay time $\tau_1= 8$ ns and try different values of $a_1$, as shown in Fig.~\ref{sfig: pre-dist calib}(d). For each value of $a_1$, we evaluate a cost function $c(a_1) = \sum_i \varphi_\mathrm{s}(a_1, t_{\mathrm{d}, i})^2$ based on the axis shift values measured for different delays $\{t_{\mathrm{d}, i}\}$. Sweeping $a_1$ and trying few different pulse durations, we find that the cost function reaches its minimum at approximately $a_1=-0.028$, where its value is an order of magnitude smaller compared to $a_1=0$ for pulse durations $t_\mathrm{p} \sim 10$ ns. Using the obtained parameters $\tau_1=8$ ns and $a_1=-0.028$, we illustrate the in-phase envelope of a FAST DRAG-P pulse  with and without predistortion in Fig.~\ref{sfig: pre-dist calib}(f).

\begin{figure*}[t!]
\centering
    \includegraphics[width = 0.9\textwidth]{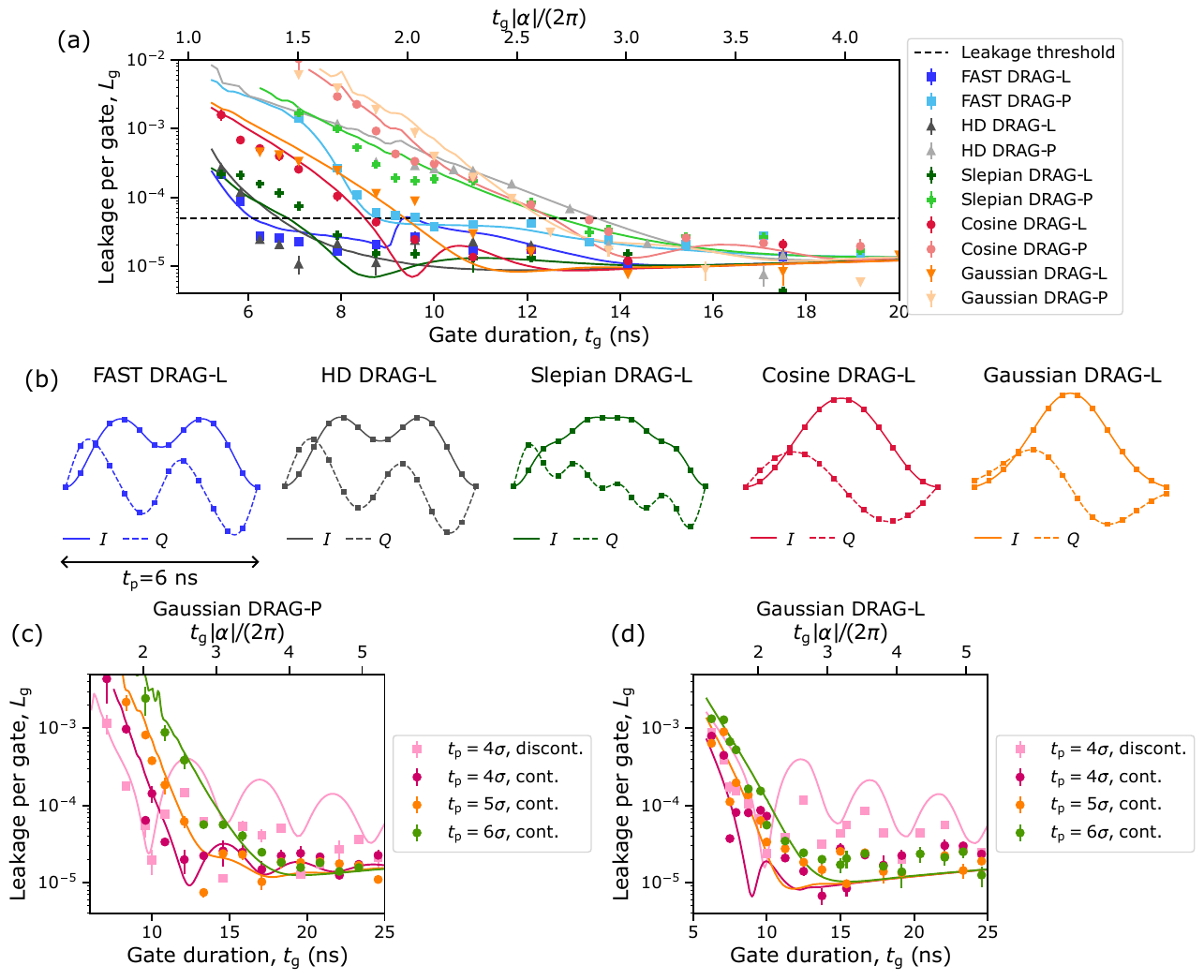}
    \caption{ \label{sfig: leakage for all pulses} \justifying  \textbf{Relation between leakage and gate duration.} (a) Experimental leakage error per gate $L_\mathrm{g}$ (markers) and simulated leakage error of $R_X(\pi/2)$ as functions of gate duration ($t_\mathrm{g} = t_\mathrm{p} + 0.41$ ns) for pulse shapes based on FAST DRAG (blue), HD DRAG (grey), Slepian DRAG (green), Cosine DRAG (red), and Gaussian DRAG (orange) calibrated using DRAG-P (light color) or DRAG-L (dark color). The experimental results are based on the mean of 3-14 RB measurements with the error bars showing the standard error of the mean. For Gaussian DRAG, we use $\sigma = t_\mathrm{p} / 5$ and subtract the discontinuity of the in-phase envelope. For Slepian DRAG, we minimize the spectral energy above $f_\mathrm{c} = 185$ MHz. 
    The dashed line shows the leakage threshold of $L_\mathrm{g}=5 \times 10^{-5}$ used to determine the speed limit for Fig. \ref{fig: stability and comparison}(a). 
    (b)~In-phase ($I$, solid) and quadrature ($Q$, dashed) envelopes for the pulse shapes of panel (a) using DRAG-L and a pulse duration of $t_\mathrm{p}=6$ ns. Square markers illustrate the sampling rate of our AWG.
    (c)~Experimental leakage per gate (markers) and simulated leakage error for $R_X(\pi/2)$ as functions of the gate duration for different versions of Gaussian DRAG-P. For the different versions, we vary the $t_\mathrm{p}/\sigma$-ratio  using values from the set $t_\mathrm{p}/\sigma \in \{4, 5, 6\}$. We also either subtract the offset of the Gaussian making the in-phase envelope continuous (dark color) or do not subtract the offset leaving the in-phase envelope discontinuous (light color). 
    (d) Same as (c) but calibrating the DRAG coefficient using DRAG-L. }
\end{figure*}

We point out that the current calibration procedure for the predistortion parameters $\{\tau_1, a_1\}$ is relatively time-consuming, but fortunately the same calibration works across different pulse shapes and gate durations. We observed that the calibrated parameters systematically improved the gate performance over the cooldown, during which the data for the main text was collected. In a subsequent cooldown without any changes to the cabling, the I-distortions measured from the I-distortion characeterization experiment were observed to no longer follow an exponential decay and could not be corrected well using an exponential IIR inverse filter with a single exponential term. After a further thermal cycle without any changes to the cabling, the simple exponential decay was recovered in the I-distortion characterization experiment and essentially the same parameters $\tau_1=8$ ns and $a_1=-0.028$ were found to provide optimal performance. 

We also attempted to correct the C-distortions observed in  Fig.~\ref{fig: MW distortions}(f) by assuming the C-distortions to be described by a cross-quadrature exponential IIR filter model of the form 
\begin{equation}
    y(t) = u(t) \times (1 + \mathrm{i} \sum_j \tilde{a}_j \mathrm{e}^{-t/\tilde{\tau}_j}),  \label{seq: cross-quadrature exponential IIR filter}
\end{equation}
where $y(t)=y_\mathrm{I}(t) + \mathrm{i} y_\mathrm{Q}(t)$ is the complex envelope seen by the qubit, $u(t) = u_\mathrm{I}(t) + \mathrm{i}u_\mathrm{Q}(t)$ is the step function in both $I$ and $Q$ components, and $\{\tilde{a}_j,\tilde{\tau}_j\}$ are the distortion parameters. However, we were not able reduce coherent errors observed in the C-distortion characterization experiment by predistorting the control pulses under the distortion model of Eq.~\eqref{seq: cross-quadrature exponential IIR filter} despite optimizing the distortion parameters. Hence, a more appropriate model for the C-distortions would be needed to enable successful predistortion since  our distortion characterization experiments require the assumption of a specific filter model. Therefore, further research is needed to understand the exact origins of the distortions and to devise more appropriate filter models. One promising distortion model to be tested in the future are control line reflections that could result in both I- and C-distortions due to an additional phase accrued by the reflected pulses. An alternative direction of research is to develop methods for characterizing the full transfer function of the control lines, e.g., by extending the methods of Ref.~\cite{jerger2019situ}.

\section{\label{ap: further leakage results} Further results on leakage error for different pulse shapes}
\begin{figure*}[t]
\centering
    \includegraphics[width = 0.92\textwidth]{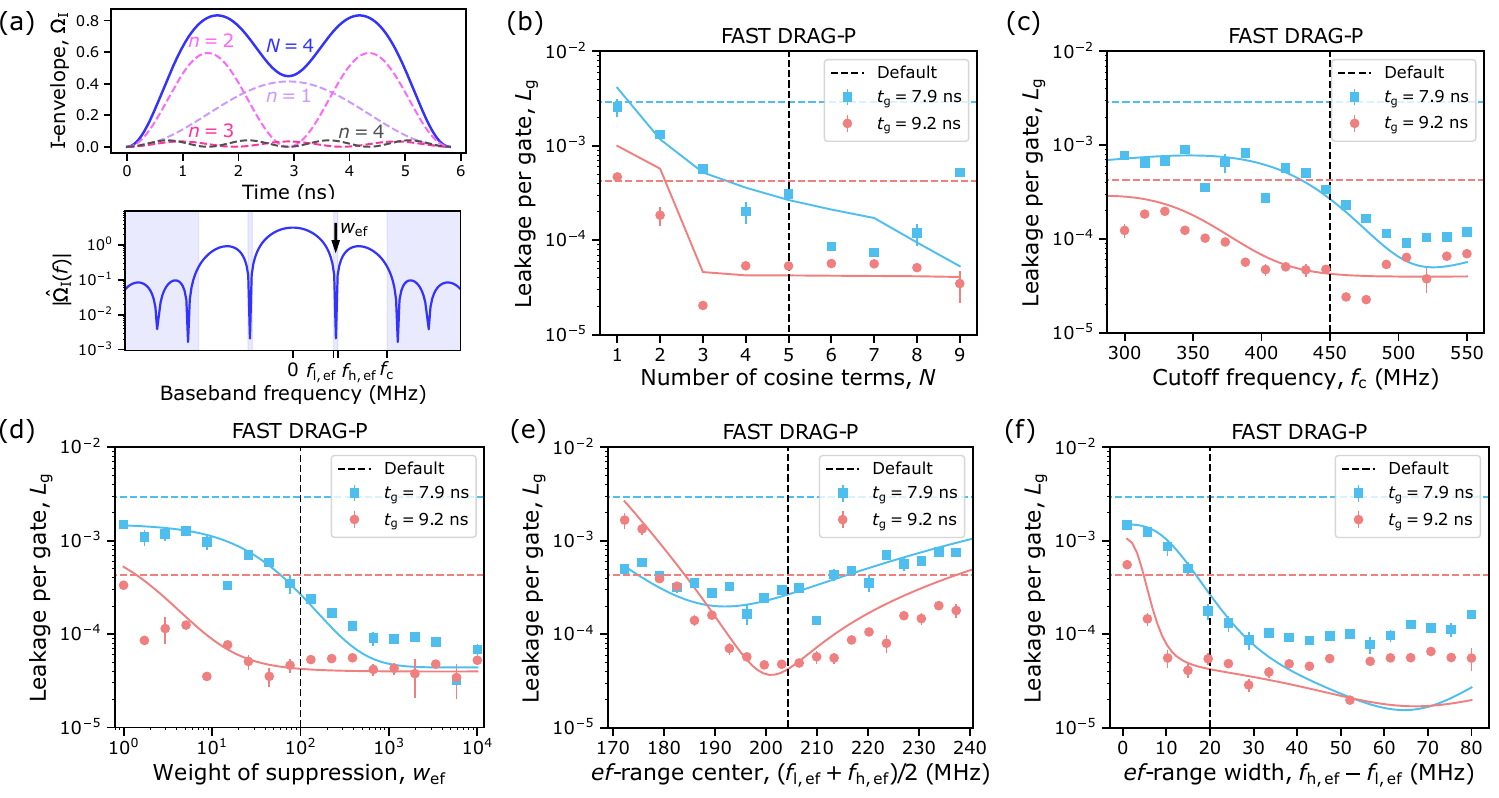}
    \caption{ \label{sfig: FAST DRAG-P sweeps} \justifying \textbf{Dependence of leakage error on the parameters of FAST DRAG-P pulses.}  (a) Schematic of the FAST DRAG parameters varied in panels (b)-(f). The top figure illustrates the use of $N=4$ cosine waves to construct the $I$-envelope, whereas the lower figure shows the amplitude spectrum $|\hat{\Omega}_\mathrm{I}(f)|$ of the $I$-envelope together with parameters controlling it, including the weight ratio $w_\mathrm{ef} = w_1/w_2$, the frequency interval suppressing the $ef$-transition $[f_{\mathrm{l}, \mathrm{ef}}, f_{\mathrm{h}, \mathrm{ef}}]$, and  the cutoff frequency $f_\mathrm{c}$ controlling the bandwidth of the pulse. The shaded blue regions show the resulting suppressed frequency intervals. (b) Experimental leakage error per gate $L_\mathrm{g}$ (markers)  and simulated leakage error of $R_X(\pi/2)$ (solid lines) as functions of the number of cosine terms $N$ for gate durations ($t_\mathrm{g} = t_\mathrm{p} +  0.41$ ns) of $t_\mathrm{g} = 7.9$ ns (blue) and $t_\mathrm{g} = 9.2$ ns (red). The experimental results are based on the mean of 3-4 repeated RB experiments while the error bars show the standard error of the mean.
    The dashed colored horizontal lines correspond to the experimental leakage error of Cosine DRAG-P  for the corresponding gate durations, 
    whereas the vertical dashed line shows the default parameter value used in the main text. (c) Same as (b) but as a function of the cutoff frequency $f_\mathrm{c}$. (d) Same as (b) but as a function of the weight ratio $w_\mathrm{ef}$. (e) Same as (b) but as a function of the center of the suppressed frequency interval $(f_{\mathrm{l}, \mathrm{ef}} + f_{\mathrm{h}, \mathrm{ef}}) / 2$, while fixing the width of the suppressed frequency interval to the default value of 20 MHz. (f) Same as (b) but as a function of the width of the suppressed frequency interval $f_{\mathrm{h}, \mathrm{ef}} - f_{\mathrm{l}, \mathrm{ef}}$, while keeping the center of the suppressed frequency interval at the default value of 204 MHz. The data for this figure were collected in a later cooldown compared to the results in the main text. }
\end{figure*}

Here, we compare experimental and simulated leakage error as a function of gate duration, see Fig.~\ref{sfig: leakage for all pulses}(a), for all the studied pulse shapes: FAST DRAG-L/P, HD DRAG-L/P, Cosine DRAG-L/P, Slepian DRAG-L/P and Gaussian DRAG-L/P. In general, the experimental results agree with the simulations across the different pulse shapes and gate durations. As an exception, the simulated leakage rate for Slepian DRAG-L  is lower than measured for the fastest gates as discussed in Appendix~\ref{ap: FAST param sweeps}.
To obtain the speed limit  presented in Fig.~\ref{fig: stability and comparison}(a) for each of the pulses, we linearly interpolate the experimental leakage error to estimate the gate duration, for which the leakage error exceeds the threshold value of $L_\mathrm{g} = 5\times 10^{-5}$. 
From Fig.~\ref{sfig: leakage for all pulses}(b) showing the envelopes of the studied pulses, we see that the envelopes of FAST DRAG-L and HD DRAG-L resemble each other for the used parameters and require a lower peak amplitude compared to Cosine DRAG-L or Gaussian DRAG-L.

We further study the leakage of the conventional Gaussian DRAG pulses by varying gate duration, $t_\mathrm{p}/\sigma$-ratio, and removal of the offset $\mathrm{exp}[-t_\mathrm{p}^2/(8\sigma^2)]$ resulting in a continuous or discontinuous in-phase envelope. We conclude that Gaussian DRAG does not enable as fast gates with low leakage as does FAST DRAG or HD DRAG for any set of parameters, see Figs. \ref{sfig: leakage for all pulses}(c) and (d) for DRAG-P and DRAG-L.
The pulse with the discontinuous Gaussian envelope suffers from a significantly higher leakage error for $t_\mathrm{g} > 15$ ns compared to the continuous envelopes. However, the experimental leakage error is somewhat lower than the result obtained from simulations, in which we ignore any non-idealities of the control electronics, see Appendix~\ref{ap: 1qb gate simulations}. Thus, we suspect that the finite bandwidth of the physical control electronics effectively reduces the abrupt in-phase discontinuity and the spectral power at the $ef$-transition, resulting in a lower leakage in the experiments.

\begin{figure*}[t]
\centering
    \includegraphics[width = 0.92\textwidth]{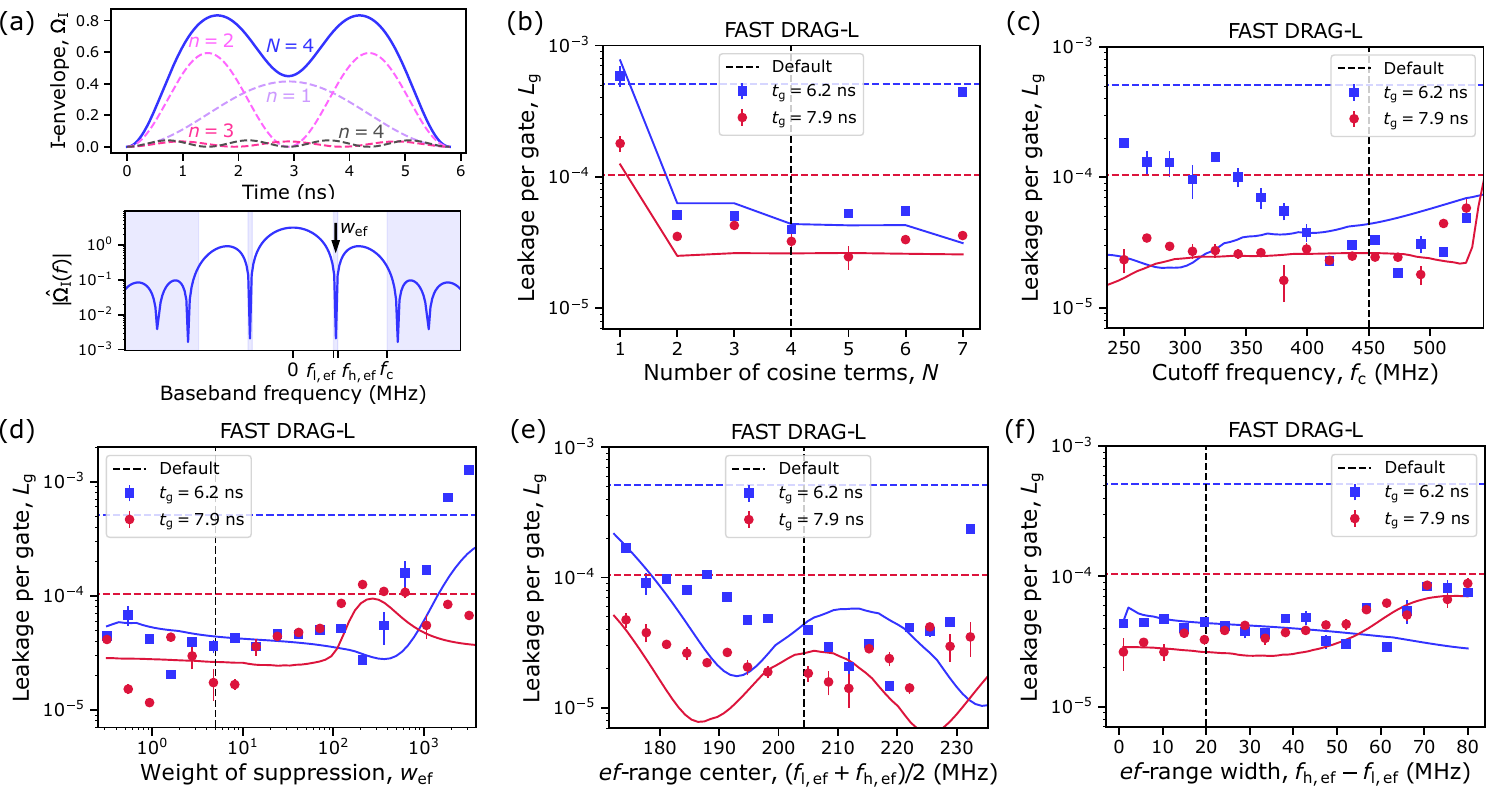}
    \caption{ \label{sfig: FAST DRAG-L sweeps} \justifying \textbf{Dependence of leakage error on the parameters of FAST DRAG-L pulses.} (a) Schematic of the FAST DRAG parameters varied in panels (b)-(f). See also the caption of Fig.~\ref{sfig: FAST DRAG-L sweeps}(a) for more detailed explanation. (b) Experimental leakage error per gate $L_\mathrm{g}$ (markers) and simulated leakage error for $R_X(\pi/2)$ (solid lines) as a function of the number of cosine terms $N$ for gate durations ($t_\mathrm{g} = t_\mathrm{p} +  0.41$ ns) of $t_\mathrm{g} = 6.2$ ns (blue) and $t_\mathrm{g} = 7.9$ ns (red). The experimental results are based on the mean of 4 repeated RB experiments while the error bars show the standard error of the mean. 
    The dashed colored horizontal lines show the experimental leakage rate of the Cosine DRAG-L pulse for the corresponding gate durations, 
    whereas the vertical dashed line shows the default parameter value used in the main text. (c) Same as (b) but as a function of the cutoff frequency $f_\mathrm{c}$. (d) Same as (b) but as a function of the weight ratio $w_\mathrm{ef}$. (e) Same as (b) but as a function of the center of the suppressed frequency interval $(f_{\mathrm{l}, \mathrm{ef}} + f_{\mathrm{h}, \mathrm{ef}}) / 2$, while fixing the width of the suppressed frequency interval to the default value of 20 MHz. (f) Same as (b) but as a function of the width of the suppressed frequency interval $f_{\mathrm{h}, \mathrm{ef}} - f_{\mathrm{l}, \mathrm{ef}}$, while keeping the center of the suppressed frequency interval at the default value of approximately 204 MHz. The data for this figure were collected in a later cooldown compared to the results in the main text.}
\end{figure*}

Furthermore, we see from  Figs. \ref{sfig: leakage for all pulses}(c) and (d) that the experiments replicate relatively well the narrow minima in leakage error  especially for the discontinuous Gaussian envelope with $t_\mathrm{p}/\sigma=4$. The narrow leakage minima are caused by one of the pulse-duration-dependent dips of the in-phase spectrum coinciding  with the $ef$-transition, see the analogous in-phase spectrum of the cosine envelope in Fig.~\ref{fig: intro and FAST}(c). Since the location of these spectral minima are dependent on the pulse duration for the conventional pulse shapes, they do not enable reaching a low leakage rate across a range of gate durations. As a final observation from Figs.~\ref{sfig: leakage for all pulses}(c) and (d), we see that the leakage rate is typically lower for a smaller $t_\mathrm{p}/\sigma$-ratio at short gate durations, whereas the situation may be the opposite for longer gate durations. 
We find that the continuous Gaussian envelope with $t_\mathrm{p}/\sigma=5$ provides a good compromise across the studied Gaussian parameters, though we point out that FAST DRAG, HD DRAG and Slepian DRAG provide superior performance for fast gates.

\section{\label{ap: FAST param sweeps} Parameter sweeps for FAST DRAG, HD DRAG and Slepian DRAG}

Here, we study the experimental and simulated leakage error when sweeping the parameters of FAST DRAG-L/P, HD DRAG-L, and Slepian DRAG-L. All the experimental results presented in this section have been acquired in a later cooldown but using the same qubit device and setup compared to the results of the main text.  For FAST DRAG-P/L, we consider sweeping the number of cosine terms $N$, the cutoff frequency $f_\mathrm{c}$, the weight (ratio) $w_\mathrm{ef} = w_1/w_2$ of $ef$-transition suppression, as well as the center frequency $(f_{\mathrm{l}, \mathrm{ef}} +f_{\mathrm{h}, \mathrm{ef}})/2 $ and width $f_{\mathrm{h}, \mathrm{ef}} - f_{\mathrm{l}, \mathrm{ef}} $ of the  $ef$-transition suppression. The meaning of each parameter value is further illustrated in Fig.~\ref{sfig: FAST DRAG-P sweeps}(a). We sweep only one of the parameters at a time and use the values provided in Sec.~\ref{sec: exp results} as defaults.

The parameter sweeps for the measured and simulated leakage error obtained using a FAST DRAG-P pulse are shown in  Figs.~\ref{sfig: FAST DRAG-P sweeps}(b)-(f) for gate durations $t_\mathrm{g} = t_\mathrm{p} + 0.41$ ns of $t_\mathrm{g}=7.9$ ns and $t_\mathrm{g}=9.2$ ns.
Overall, we observe that the experimental results and simulations agree well, and FAST DRAG-P reduces the leakage rate by up to a factor of 30 compared to Cosine DRAG-P (dashed horizontal lines) at $t_\mathrm{g}=7.9$ ns. From  Fig.~\ref{sfig: FAST DRAG-P sweeps}(b), 
we see that 
three cosine terms is sufficient to obtain a low leakage error at $t_\mathrm{g}=9.2$ ns,  
whereas the leakage rate is reduced  up to $N=7$ for $t_\mathrm{g}=7.9$ ns. For $N>7$, the experimental leakage rate is increased due to the emergence of high-frequency components, which we attribute to using a fixed upper limit $f_{\mathrm{h}, 2}=1000$ MHz for the  second frequency interval $[f_\mathrm{c}, f_{\mathrm{h}, 2}]$ to be suppressed. By using a higher value of $f_{\mathrm{h}, 2}$, the high-frequency components could be suppressed and this issue could be mitigated.  

\begin{figure*}[t]
\centering
    \includegraphics[width = 0.67\textwidth]{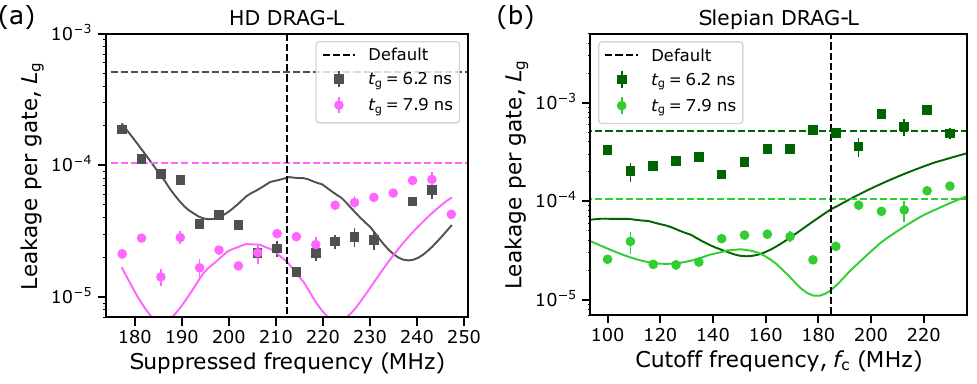}
    \caption{ \label{sfig: 3rd-der DRAG sweeps} \justifying \textbf{Dependence of leakage error on the parameters of HD DRAG-L and Slepian DRAG-L.} (a) Experimental leakage error per gate (markers) and simulated leakage error of $R_X(\pi/2)$ (solid lines)  as functions of the suppressed baseband frequency $f=1/(2\pi\sqrt{\beta_2})$ in the in-phase spectrum  of a HD DRAG-L pulse  when using a gate duration ($t_\mathrm{g} = t_\mathrm{p} +  0.41$ ns) of $t_\mathrm{g} = 6.2$ ns (grey) and $t_\mathrm{g} = 7.9$ ns (pink). The experimental results are based on the mean of 4 repeated RB experiments while the error bars show the standard error of the mean. 
    The dashed colored horizontal lines correspond to the experimental leakage error of Cosine DRAG-L pulse for the corresponding gate durations, 
    whereas the dashed vertical line shows the default parameter value used in the main text. (b) Experimental leakage error per gate (markers) and simulated leakage error of $R_X(\pi/2)$ (solid lines) as functions of the  cutoff frequency $f_\mathrm{c}$ for a Slepian DRAG-L pulse when using a gate duration ($t_\mathrm{g} = t_\mathrm{p} +  0.41$ ns) of $t_\mathrm{g} = 6.2$ ns (dark green) and $t_\mathrm{g} = 7.9$ ns (light green). The experimental results are based on the mean of 3 repeated RB experiments while the error bars show the standard error of the mean. The dashed and solid lines have corresponding meaning as in (a). The data for this figure were collected in a later cooldown compared to the results in the main text.}
\end{figure*}

From Figs.~\ref{sfig: FAST DRAG-P sweeps}(c), (d), and (f), we see that the leakage rate of FAST DRAG-P is in general reduced as the $ef$-transition is more strongly suppressed, which can be achieved by increasing the cutoff frequency $f_\mathrm{c}$, the weight ratio $w_\mathrm{ef}$ or the width $f_{\mathrm{h}, \mathrm{ef}} - f_{\mathrm{l}, \mathrm{ef}} $ of the frequency interval to be suppressed. Importantly, a stronger suppression is required to reduce the leakage rate at the shorter gate duration of $t_\mathrm{g}=7.9$ ns. We further observe that the simulations predict a somewhat lower leakage error than measured experimentally for large values of $f_\mathrm{c}$, $w_\mathrm{ef}$ and $f_{\mathrm{h}, \mathrm{ef}} - f_{\mathrm{l}, \mathrm{ef}}$. 
We suspect that the difference is caused by the simulations not accounting for the non-idealities of the control electronics, such as finite bandwidth and sampling rate that hinder the experimental performance. From Fig.~\ref{sfig: FAST DRAG-P sweeps}(e), we further see that the central frequency of the suppressed interval is slightly shifted when changing the gate duration, though the minimum is relatively broad.

Figure~\ref{sfig: FAST DRAG-L sweeps} shows the  experimental and simulated parameter sweeps of the leakage error for FAST DRAG-L at gate durations $t_\mathrm{g} = t_\mathrm{p} + 0.41$ ns of $t_\mathrm{g}=6.2$ ns and $t_\mathrm{g}=7.9$ ns. From Fig.~\ref{sfig: FAST DRAG-L sweeps}(b), we see that using only $N=2$ cosine terms results in a low leakage error even for the gate duration of $t_\mathrm{g} = 6.2$ ns.  Based on Figs.~\ref{sfig: FAST DRAG-L sweeps}(c), (d), and (f), FAST DRAG-L achieves a low leakage error with less aggressive suppression of the in-phase spectrum compared to FAST DRAG-P since the DRAG coefficient $\beta$ also contributes to leakage minimization. Hence, a low leakage rate is reached at significantly lower values of $f_\mathrm{c}$, $w_\mathrm{ef}$ and $f_{\mathrm{h}, \mathrm{ef}} - f_{\mathrm{l}, \mathrm{ef}}$ compared to FAST DRAG-P, and in fact, too high values of these parameters increase leakage. From Figs.~\ref{sfig: FAST DRAG-L sweeps}(b)-(f), we further observe that moderate changes to the parameter values have a relatively small effect on the leakage error, which can be regarded as a sign of robustness. We observe that the differences between the simulated and measured leakage rates are higher compared to FAST DRAG-P, which is likely caused by the use of shorter gate durations that may be more susceptible to the non-idealities of the control electronics.

From Fig.~\ref{sfig: 3rd-der DRAG sweeps}(a) showing the leakage error of HD DRAG-L as a function of the suppressed frequency, we observe that the minimum of the leakage error occurs approximately at a suppressed (baseband) frequency of $|\alpha|/(2\pi)$ corresponding to the default value used to obtain the results of the main text. Furthermore, we find that the minimum is relatively broad. When sweeping the cutoff frequency of Slepian DRAG-L as shown in Fig.~\ref{sfig: 3rd-der DRAG sweeps}(b), we observe that the default cutoff frequency of $f_\mathrm{c}=185$ MHz provides a representative leakage error that cannot be much improved by further optimizing the cutoff frequency. We also observe that the measured leakage rate at $t_\mathrm{g} = 6.25$ ns is systematically higher than the simulated leakage, which may again be caused by the finite sampling rate and bandwidth of the control electronics that do not capture well the rapid changes in the quadrature envelope of the Slepian DRAG-L pulse, see Fig.~\ref{sfig: leakage for all pulses}(b). 

\pagebreak

%\bibliography{bibliography}% Produces the bibliography via BibTeX.
\bibliography{main}

\iffalse


\fi

\end{document}